\documentclass{article}
 \newcommand{\ads}{(a^{\dag})^{s}}
 \newcommand{\bdt}{(b^{\dag})^{t}}
 \newcommand{\ad}{a^{\dag}}
 \newcommand{\p}{\prime}
 \newcommand{\adsp}{(a^{\dag})^{s^{\p}}}
 \newcommand{\adspp}{(a^{\dag})^{s^{\p\p}}}
 \newcommand{\bdtp}{(b^{\dag})^{t^{\p}}}
 \newcommand{\bdtpp}{(b^{\dag})^{t^{\p\p}}}
 \newcommand{\bd}{b^{\dag}}
 \newcommand{\dd}{d^{\dag}}
 \newcommand{\cd}{c^{\dag}}
 \newcommand{\xd}{x^{\dag}}
 \newcommand{\g}{\gamma}
 \newcommand{\gp}{\gamma^{\prime}}

 \newcommand{\asbt}{\g,s;\gp,t}
 \newcommand{\asmbt}{\g,(m,h),\ads;\gp,\bdt}
 \newcommand{\asbtp}{\g_{1}s';\gp_{1},t'}
 \newcommand{\asmbtp}{\g_{1},(m,h),\adsp;\gp_{1},\bdtp}
 \newcommand{\asbtpp}{\g'_{1},s";\g"_{1},t"}
 \newcommand{\ap}{\alpha^{\p}}
 \renewcommand{\sp}{s^{\p}}
 \newcommand{\bp}{\beta^{\p}}
 \newcommand{\tp}{t^{\p}}
 \renewcommand{\a}{\alpha}
 \renewcommand{\b}{\beta}
 
           \usepackage{graphics}
          \begin{document}
           \title{A Representation of Real and Complex Numbers in
           Quantum Theory}
          \author{Paul Benioff\\
           Physics Division, Argonne National Laboratory \\
           Argonne, IL 60439 \\
           e-mail: pbenioff@anl.gov}
           \date{\today}
           \maketitle

          \begin{abstract}
          A quantum theoretic representation of real and complex numbers
          is described here as equivalence classes of Cauchy sequences
          of quantum states of finite strings of qubits. There are $4$
          types of qubits each with associated single qubit annihilation
          creation (a-c) operators that give the state and location of each
          qubit type on a $2$ dimensional integer lattice. The string
          states, defined as finite products of creation operators acting
          on the vacuum state $|0\rangle,$ correspond to complex rational
          numbers with real and imaginary components. These states span a
          Fock space $\mathcal F.$ Arithmetic relations and operations are
          defined for the string states.  Cauchy sequences of these states
          are defined, and the arithmetic relations and operations lifted to
          apply to these sequences. Based on these, equivalence classes of
          these sequences are seen to have the requisite properties of
          real and complex numbers. The representations have some
          interesting aspects. Quantum equivalence classes are larger
          than their corresponding classical classes, but no new classes
          are created. There exist Cauchy sequences  such that each
          state in the sequence is an entangled superposition of the
          real and imaginary components, yet the sequence is a real
          number. Also, except for coefficients of
          superposition states, the construction is done with no reference
          to the real and complex number base, $R,$ $C,$ of $\mathcal F$
          \end{abstract}

          \section{Introduction}
          Real and complex numbers are very important to physics
          in several different ways. They form the basis of all
          physical theories in that the theories are mathematical structures
          based on the  real and complex numbers. Real numbers are also
          used to represent the space time manifold as $R^{4}$. All
          theoretical predictions to be tested by experiment are, or
          can be cast, as real number solutions to equations.

          On the other hand outputs of experiments are rational
          numbers.  This is based on the observation that they are
          or can be represented as  states of finite strings of kits
          or qukits in some base $k\geq 2.$ Also all computers, both
          classical and quantum, work with states of finite strings
          of kits or qukits. Usually they are base $2$ rational
          numbers as states of finite bit or qubit strings.

          Comparison between experiment outputs and computer
          outputs as a comparison between theory and experiment
          depends on the fact that rational numbers are dense in
          the set of real numbers. Also those rational numbers
          expressible as states of finite base $k$ kit or qukit
          strings, are dense in the set of all rational numbers.
          Because of this computer outputs, as states of finite
          kit or qukit strings, represent, to arbitrary accuracy,
          theoretical predictions. Also they can be
          directly compared to experimental predictions.

          However, as noted, physical theories are based on real and
          complex numbers and not on rational numbers.  The
          completeness properties of real and complex numbers
          play an essential role in theoretical
          predictions.  This follows from the observation that all
          theoretical predictions are theorems, i.e. theoretical
          statements in the theory language that are provable in
          the  physical theory.  The proofs of these statements,
          which are based on the mathematical properties of the
          theory as an axiomatizable mathematical theory,
          depend essentially on the properties of the real and
          complex number base of the theory.

          These considerations show the basic importance of the
          different types of numbers to physics and mathematics. Yet
          they leave open the deeper question
          of the relationship between the foundations of mathematics
          and physics and why mathematics is relevant to physics.
          This question, which  was first described by \cite{Wigner}
          and commented on by others \cite{Hamming,Davies}, is
          especially acute if one accepts the Platonic view of
          mathematical existence.  In this view, which seems to be
          accepted at least implicitly by many, mathematical objects
          have an ideal abstract existence.  This seems completely
          unrelated to the physical existence of objects that both exist
          in and determine the properties of space time.

          There are several different approaches to understanding this
          relationship \cite{Tegmark1}-\cite{Weinberg}. The approach
          underlying this paper  is to work towards a coherent theory of
          physics and mathematics together \cite{BenTCTPM}. Such a theory,
          by treating both physics and mathematics together in one theory,
          may help to understand how mathematics and physics are related.
          It may also help to answer some of the basic outstanding questions
          in physics.

          In this paper a step in this direction is taken by
          describing quantum representations of real and complex
          numbers. The use of quantum rather than classical
          representations is done because this brings both the
          treatment of physical systems and numbers into the same
          general theory. Quantum theory is the basic theory
          underlying the description of physical systems. Using
          the same basic theory to describe both physical systems
          and numbers as mathematical systems should help in
          bringing together descriptions of physical and
          mathematical systems.(The question of the relevance of
          real and complex numbers in physics \cite{Isham} will
          not be treated here.)

          The other main point is that all physical representations
          of numbers are as states of finite strings of physical
          systems.  This will be taken over here in that the only
          systems available for representations of numbers are
          states of finite strings of kits or qukits.  Since quantum
          theory is taken to be the basic underlying theory for both
          physical and mathematical systems, quantum representations
          of real and complex numbers will be based on states of finite
          strings of qukits.

          The importance of qukits lies in the fact that they are
          the basic units of quantum information just as kits are
          the basic units of classical information. The importance
          is based on the observation that qukits, as units with
          $k$ orthogonal  states for any $k>2,$ can be used for
          either quantum representations of numbers, or for
          representations of quantum mechanical systems in physics.

          States of finite strings of qukits are quite useful to
          represent the natural numbers $N$, the integers, $I,$ and
          the rational numbers $Ra.$  However they do not represent
          either real or complex numbers\footnote{They can be
          represented by states of infinte strings of qukits, but
          these are not describable as states in a separable Hilbert
          space.  Even a field theoretic description seems problematic
          even though systems with an infinite number of degrees of
          freedom are described in quantum field theory.}. Thus some
          way to connect these representations of rational numbers
          to real and complex numbers must be found.

          The method used here follows the one in some
          mathematical analysis textbooks \cite{Burkill} that describe
          real numbers as equivalence classes of Cauchy sequences
          of rational numbers. The other equivalent description as
          equivalence classes of Dedekind cuts is not used here.
          Also complex numbers are usually described as ordered pairs
          $(x,y)$ (or $(x,iy)$) of real numbers.  However here, the
          procedure used in computers that work with complex rational
          numbers as ordered pairs of rational numbers, $(ra_{1}, ira_{2})$
          will be followed.

          The goal of this paper is to give quantum theory
          representations of real and complex numbers as equivalence
          classes of Cauchy sequences of quantum representations of
          real and complex rational numbers as states of finite
          strings of qukits. To avoid inessential complications, the
          description will be limited  $k=2$ or states of strings
          of qubits.

          The use of quantum theory to study representations
           of numbers and other mathematical systems is not new
           \cite{Litvinov} -\cite{Krol}. Of particular note is
           work on quantum set theory represented as an orthomodular
           lattice valued set theory \cite{Takeuti}-\cite{Krol}. In
           this work natural numbers, integers, and rational numbers
           have  representations that are either similar to the usual
           ones in mathematical analysis \cite{Takeuti}-\cite{Titani}
           or are based on a categorical approach \cite{Schlesinger,Krol}.
           However the work in these references differs from the approach
           taken here in that, here, states of finite qubit strings are used as
           the rational number base of real and complex numbers.

          Steps taken in this paper in this include, descriptions of  rational
          numbers as states of finite strings of qubits by use of annihilation
          creation operator strings acting on the qubit vacuum state, and
          the description of the basic arithmetic relations and operations on
          these rational string states. This is done in the next section.

          Cauchy sequences of these states are defined in Section \ref{CSSF}.
          The definitions are based on the description of the arithmetic
          relations and operations. Some examples are given including some
          that have no classical counterpart. The definition of Cauchy
          convergence is also extended to sequences of states that are
          linear superpositions of rational string states.

          Section \ref{POCSS} describes the basic properties and operations
          on Cauchy sequences.  This includes lifting of the basic
          arithmetic relations and operations to apply to the sequences.
          Section \ref{RRCNQT} describes the real and complex number
          representations as equivalence classes of the Cauchy sequences.
          The proof that they, or equivalently, representatives of each class
          satisfy the real number axioms is outlined. It is seen
          that the quantum equivalence classes are larger than the
          classical ones but that no new classes are created.

          It is useful to note terminology used in the paper.
          Following standard usage real, imaginary, and complex real
          numbers will be referred to as real, imaginary, and
          complex numbers.  Real, imaginary and complex rational
          numbers will be referred to as noted.  Rational string
          states are states of finite qukit strings that represent
          rational numbers. Complex rational string states are
          states of pairs of finite qukit strings that represent
          complex rational numbers.

          The last section contains a discussion of these results.
          It is emphasized again that, except for the description of
          linear superpositions of string states and Cauchy sequences
          of these states, the quantum theoretic description of real and
          complex numbers is independent of $R$ and $C$. It is also noted
          that, since the equivalence classes of Cauchy sequences of
          quantum states are complex numbers, they can be
          used as the complex number field for any physical theory
          that is a mathematical structure based on the complex
          number field.  This encompasses most of physics, since
          theories such as quantum mechanics, QED, QCD, string
          theory, and general relativity are structures of this type.
          The section also has a brief discussion of the relation
          between this complex number field and $C$.

          \section{Complex Rational String States}\label{CRSS}

          It is useful to define states of pairs of qubit strings on
          a two dimensional integer lattice $I\times I.$ One
          dimension indicates the string direction and the other
          allows for an arbitrary number of strings as these are
          needed for the discussion of $n-ary$ operations on the
          strings.

          For this paper it should be noted that it is not necessary
          to consider $I\times I$ as a lattice of points in $2$ or
          more dimensional physical space. This would be suitable for
          physical representations of the mathematical model being
          considered here.  Here the lattice $I\times I$ is a
          convenient method to represent the minimal conditions
          needed here.  These are \begin{itemize}
          \item $I$ is an infinite set of points  with the order
          type of the integers: \begin{enumerate}\item No largest
          or smallest point, \item Ordering is discrete, \item Each
          point has just one nearest neighbor above and below.
          \end{enumerate}\item A denumerable set of pairs of discrete points
          whose purpose is to distinguish the qubits in different
          strings. This is especially important for fermionic
          qubits. \item No metric distance between pairs of
          lattice points is assumed or needed here.\end{itemize}
          Note that ordering of the labels of the different strings
          is convenient, and is represented by the second $I$
          component of the lattice. But it is not necessary.

          For the purposes of this paper it is immaterial whether
          the string pairs needed to represent a complex rational
          number consist of one string of two different types of
          qubits or two adjacent strings of the same type of qubit
          that are distinguished by their different positions on the
          lattice.  Here the model consisting of one string of two
          different qubit types will be used.

           A compact representation of numbers is used here that combines
          the location of the sign and the "binal" point. The representation
          is suitable for  real and complex natural numbers and
          integers as well as rational numbers. As examples the
          usual decimal numbers, $-63.71,\; 459,\; -0.0753$
          would be expressed here as $63-71,\; 459+,\;0-0753$.

          Since strings with different numbers of qubits are being
          considered, it is useful to describe them using
          annihilation creation (AC) operators. Here four types are
          needed: $\ad_{\a,j,h},a_{\a,j,h}; \bd_{\b,j,h},b_{\b,j,h};
          \cd_{\g,j,h}c_{\g,j,h};\dd_{\g,j,h},d_{\g,j,h}.$ The
          $c$ and $d$ qubits denote the signs of the real and
          imaginary strings where $\g =+,-$, and the $a$ and
          $b$ qubits are the $0,1$ components of the real and
          imaginary qubit strings. Here $\a$ and $\b$ take values
          in $\{0,1\}$ and $(j,h)$ is a point in the lattice.

          Complex rational string states are represented by
          \begin{equation}\label{comrastes}|\g,\ads;\gp,\bdt,(m,h)\rangle
          =\cd_{\g,m,h}\ads_{[l,u]};\dd_{\gp,m,h}\bdt_{[l',u']}|0\rangle
          \end{equation}where the state $|0\rangle$ denotes the qubit
          vacuum. Here $m,h$ denotes the location of the sign
          qubits, $[l,u]$ and $[l',u']$ denote lattice intervals
          $(l,h),(l+1,h),\cdots,(u,h)$ and$(l',h),(l'+1,h),\cdots,(u',h)$
          where $l\leq m\leq u$ and $l'\leq m\leq u'$.   Finally
         \begin{equation}\label{asbt}\begin{array}{c}\ads_{[l,u]}
          =\ad_{s(u),u,h}\ad_{s(u-1),u-1,h}\cdots\ad_{s(l),l,h}\\ \bdt_{[l',u']}
          =\bd_{t(u'),u',h}\bd_{t(u'-1),u'-1,h}\cdots\bd_{t(l'),(l',h)}\end{array}
         \end{equation} where $s$ and $t$ are $0,1$ valued functions
         with integer interval domains $[l,u]$ and $[l',u']$ respectively.
         It is easy to generalize to let $s$ and $t$ be functions that depend
         also on $h,$ but this will not be done here.

         Note that the states of qubit strings described here
         represent colocated strings of two types of qubits with sign
         qubits at the location $(m,h).$ Each string location $(j,h),$ other than
         $(m,h),$ contains up to two qubits, none, or one $a$ and/or one
         $b$ type. The $(m,h)$ site contains the same $a$ and $b$
         type qubits and two sign $c$ and $d$ type qubits.

         For the purposes of this work it is immaterial whether the
         qubits are bosons or fermions as the representation is
         sufficiently inclusive to incorporate both.  Fermion AC
         operators satisfy the anticommutation relations,
         \begin{equation}\label{fercomm}\begin{array}{l}
         \{a_{\a ,j,h},a_{\ap,k,h^{\p}}\}= \{b_{\b,j,h},b_{\bp,k,h^{\p}}\}
         \\ \hspace{.5cm}=\{\ad_{\a,j,h},\ad_{\ap,k,h^{\p}}\}=
         \{\bd_{\b,j,h},\bd_{\bp,k,h^{\p}}\}=0; \\ \hspace{.5cm}\mbox{$\{ a_{\a,j,h},
         \ad_{\ap,k,h^{\p}}\} = \delta_{\a,\ap}\delta_{j,k}
         \delta_{h,h^{\p}}$}; \\ \hspace{.5cm}\mbox{$\{ b_{\b,j,h},
         \bd_{\bp,k,h^{\p}}\} = \delta_{\b,\bp}\delta_{j,k}
         \delta_{h,h^{\p}}$}\end{array}\end{equation} where
         $\{x,y\}=xy+yx.$  As a consequence a specific ordering of
         the AC operators of each type must be used.  This
         will be implicitly assumed here to be that shown in Eqs.
         \ref{comrastes} and \ref{asbt}. For bosons the same
         relations hold if $\{,\}$ is replaced by commutation relations
         $[,]$ where $[x,y]=xy-yx.$  Since the $a,b,d,c$ systems are
         all different all six pairs of these AC operators commute.
         For these systems the ordering of the AC operators does not
         matter.

         The Fock space spanned by states of the form
         $|\g,\ads;\gp,\bdt,(m,h)\rangle$ is denoted by
         $\mathcal{F}_{m,h}.$ This is the space of all complex
         rational string states with the "binal" point at $(m,h).$
         The space\begin{equation}\label{globalF}
         \mathcal{F}=\bigoplus_{(m,h)\epsilon I\times I}\mathcal{F}_{m,h}
         \end{equation} is spanned by all complex\footnote{The term
         "complex" includes both real and imaginary components.}
         rational string states located anywhere on $I\times I.$

         The arithmetic and numerical properties of the states
         $|\g,(m,h),\ads;\gp,\bdt,\rangle$ will be described here
         independent of the corresponding numerical value in the
         complex number base $C$ of the space $\mathcal{F}.$
         Nevertheless it is useful to define an operator $\tilde{N}$
         whose eigenvalues correspond to the complex numbers
         associated with the states
         $|\g,(m,h),\ads;\gp,\bdt\rangle.$  In particular
         $\tilde{N}$ serves as a check on the consistency of the
         definitions of the basic arithmetic relations and
         operations.

         $\tilde{N}$ is the sum of two operators $\tilde{N}^{R},
         \tilde{N}^{I}$ for the real and imaginary component
         states. Each of the two operators is in turn a product of
         two commuting operators, a sign scale operator
         $\tilde{N}^{X}_{ss}$, and a value operator
         $\tilde{N}^{X}_{v}$ for $X=R,I.$ One has\begin{equation}
          \label{def1N}\tilde{N}=\tilde{N}^{R}+\tilde{N}^{I}\end{equation}
          where \begin{equation}\label{NXssv} \tilde{N}^{X} =
          \tilde{N}^{X}_{ss}\tilde{N}^{X}_{v}\end{equation} and
          \begin{equation}\label{NXss}
          \tilde{N}^{X}_{ss} =\left \{\begin{array}{l}\sum_{\g,m}\g
          2^{-m}\cd_{\g,m}c_{\g,m}\mbox{ if $X$ =$R$} \\\sum_{\gp,m}
          \gp 2^{-m}\dd_{\g,m}d_{\gp,m} \mbox{ if
          $X=I$}\end{array}\right.\end{equation} and \begin{equation}\label{NXv}
          \tilde{N}^{X}_{v} =\left \{\begin{array}{l}\sum_{\a,j,h}\a 2^{j}
          \ad_{\a,(j,h)}
            a_{\a,(j,h)}\mbox{ if $X$ =$R$} \\\sum_{\b,j,h}i\b 2^{j}\bd_{\b,(j,h)}
            b_{\b,(j,h)} \mbox{ if $X=I$}.\end{array}\right.\end{equation}

            Note that, because of the presence of strings of leading
            or trailing $0s,$ the eigenspaces of $\tilde{N}$ are
            infinite dimensional.  Also $\tilde{N}$ is unbounded and
            has complex eigenvalues. The eigenspace for the number $0$
            includes all states of the form $|\g,\ads;\gp,\bdt,(m,h)\rangle$
            for all $(m,h)$ where $s$ and $t$ are  constant $0$ functions on
            integer intervals that include $m$. The signs can be either $+$
            or $-$. It is useful to designate these states by the
            simple form $|+,0\rangle.$

            \subsection{Basic Arithmetic Relations}\label{BAR}
            There are two basic arithmetic relations, equality
            $=_{A}$ and ordering $\leq_{A}.$ Two states $|\asmbt\rangle$ and
            $|\asmbtp\rangle$ are arithmetically equal if the real and
            imaginary parts are the same up to leading and trailing
            $0s.$ That is \begin{equation}\label{equalA}
            |\asmbt\rangle=_{A}|\asmbtp\rangle
            \end{equation} if for all $j$ in $D(s)\bigcap D(s')$
            $s(j)=s'(j)$ and for all $j$ in $D(s)-D(s')$ , $s(j)=0$
            and for all $j$ in $D(s')-D(s)$, $s'(j)=0$ with a
            similar condition holding for $t$ and $t'.$ Here $D(s)$
            is the integer domain of $s$.  Also $\g =\g_{1}$ and
            $\gp =\gp_{1}.$

            Because one is working with quantum states with boson or
            fermion properties, the definition of such an obvious relation
            as arithmetic equality has some unexpected features. These are
            related to the fact that $A$ equality depends only on the
            properties of $s,s',t,t'$ and not on the other variables
            such as the positions of the qubit sequences in the lattice.

            To see these features, consider a  restricted definition
            of $A$ equality $=_{A,(m,h),(m',h')}$ that applies only to
            those sequence pairs with the sign qubits at $(m,h)$ and
            $(m',h')$ and is undefined elsewhere. One aspect is that the diagonal
            definition $=_{A,(m,h),(m,h)}$ is meaningless and is not
            defined anywhere. This is a result of the fact that
            for either boson or fermion qubit sequences  which
            overlap at some lattice sites, there is
            no way to determine which of the two qubits in the
            lattice overlap sites belongs to which sequence. Also
            $=_{A,(m,h),(m',h')}$ and $=_{A,(m',h'),(m,h)}$ are
            identical as they have the same domains of definition.
            Note too that $A$ equality of two states does not
            implies quantum theory equality.  Two states can be
            quite different quantum theoretically  yet be the same
            arithmetically.

            These properties of $A$ equality are mirrored in the
            properties of the associated quantum projection
            operators for these definitions. For the restricted
            $=_{A,(m,h),(m',h')}$ the associated projection operator
            is a product of projection operators for the real and imaginary
            components,\begin{equation}\label{PRI}\tilde{P}_{=_{A,(m,h),(m',h')}}=
            \tilde{P}^{R}_{=_{A,(m,h),(m',h')}}\tilde{P}^{I}_{=_{A,(m,h),(m',h')}}
            \end{equation} Here \begin{equation}\label{PRmhmphp}
            \tilde{P}^{R}_{=_{A,(m,h),(m',h')}}=\sum_{\g,s^{\neq 0}}
            \tilde{P}_{\g,(m,h),[s]}\tilde{P}_{\g,(m',h',),[s_{\Delta m}]}\end{equation}
            and a similar expression for $\tilde{P}^{I}_{=_{A,(m,h),(m',h')}}.$
            Here $s^{\neq 0}$ means that the sum is restricted to those $s$ with
            no leading or trailing $0s$ and $[s]$ denotes the set of
            all $s'$ that are equal to $s$ up to leading or trailing
            $0s$. $[s_{\Delta m}]$ is the set of all $s'$ equal to $[s_{\Delta
            m}]$ up to leading or trailing $0s$ and $s_{\Delta
            m}(j')=s(j)$ where $j'-m'=j-m$ or $j' =j+\Delta m.$

            $\tilde{P}_{\g,(m,h),[s]}$ is given by\begin{equation}\label{Pmhs}
            \tilde{P}_{\g,(m,h),[s]}=\sum_{s'\sim_{0}s}
         \tilde{P}_{|\g,(m,h),s',l,u\rangle}.\end{equation} In
         terms of A-C operators $\tilde{P}_{|\g,(m,h),s',l,u\rangle}$
         can be expressed as \begin{equation}\label{ACeq}
         \tilde{P}_{|\g,(m,h),s',l,u\rangle}=\cd_{\gamma,(m,h)}
         (\ad)^{s'}_{[(l,h),(u,h)]}c_{\gamma,(m,h)}
         a^{s'}_{[(l,h),(u,h)]}.\end{equation} Here
         $(\ad)^{s'}_{[(l,h),(u,h)]}
         =\ad_{s'(u,h),(u,h)}\cdots \ad_{s'(l,h),(l,h)}$
         with a similar expression for $a^{s'}_{[(l,h),(u,h)]}.$
         In Eq. \ref{Pmhs} the sum over $s'\sim_{0}s$ is over all $s'$
         that differ only by leading or trailing $0s.$ The
         dependence of $l$ and $u$ on $s'$ is implied.

          A global definition of $A$ equality, $=_{A},$ is defined by
            \begin{equation}\label{Glequal} =_{A}\leftrightarrow
            \exists(m,h),(m',h')[(m,h)\neq(m',h')\wedge =_{A,(m,h),(m',h')}].
            \end{equation}  That is two states are $A$ equal if and only if they
            are $=_{A,(m,h),(m',h')}$ for some pairs
            $(m,h),(m',h').$ The corresponding projection operator
            $\tilde{P}_{=,A}$ is defined by \begin{equation}\label{P=AGl}
            \tilde{P}_{=,A}=\sum_{(m,h)\neq (m',h')}\tilde{P}_{=_{A,(m,h),(m',h')}}.
            \end{equation} As seen in Eq. \ref{PRI}
            $\tilde{P}_{=,A}$ can be expressed as a product of
            projection operators for the real and imaginary
            components.

          To save on notation the pair $(m,h)$ will often be deleted
         in the following.  Thus $|\g,\ads;\gp,\bdt,(m,h)\rangle$ will
         be represented as $|\g,\ads;\gp,\bdt\rangle$ or even in the
         shorter form $|\g,s;\gp,t\rangle.$

         The ordering relation $\leq_{A}$ is defined separately for
         the real and imaginary components of the rational state
         pairs. For positive real components one has \begin{equation}\label{lteq}
         |+,(m,h),s,t\rangle \leq_{A,R}|+,(m',h'),s',t'\rangle\leftrightarrow
         s\sim_{0_{\Delta m}}s' \mbox{ or } s<_{\Delta m}s'. \end{equation}The
         relations $\sim_{0_{\Delta m}}$ and $<_{\Delta m}$ are
         defined by\begin{equation}\label{deltam}\begin{array}{l}
         s\sim_{0_{\Delta m}}s'\mbox{ if }\left\{\begin{array}{l}\forall j
         \epsilon [l,u] (s(j)=1\leftrightarrow s'(j+\Delta m)=1\\
         \mbox{and }\forall j'\epsilon [l'u'] (s'(j')=1\leftrightarrow
         s(j-\Delta m)=1)).\end{array}\right.\\ \\ s<_{\Delta m}s'\mbox{
         if }\left\{\begin{array}{l} \exists j\epsilon [l,u]\cap[l',u'](s(j)=0,
         s'(j+\Delta m)=1 \\ \mbox{and } s_{[j+1,u]}\sim_{0_{\Delta
         m}}s'_{[j+1+\Delta m,u']}).\end{array}\right.\end{array}
         \end{equation} The definitions of $s\sim_{0_{\Delta m}}s'$
         and $s<_{\Delta m}s'$ state that $s$ is equal to or less than
         $s'$ when differences in $m,m'$ are taken into account. These
         locations matter because they are used to determine the powers
         of $2$ associated with the values of $s$ and $s'$.

         The extension to zero and negative states is given by
         \begin{equation}\label{negordr} \begin{array}{c}|+,
         \bar{0},t\rangle \leq_{A,R}|+,s',t'\rangle\mbox{ for all
         $s'$}\\ |+,s,t\rangle\leq_{A,R}|+,s',t'\rangle
         \rightarrow |-,s',t\rangle \leq_{A,R}|-,s,t'\rangle.
         \end{array}\end{equation}

         Eq. \ref{negordr} holds for any pair $(m,h),(m',h')$ where
         $(m,h)\neq(m',h').$ As was the case for $A$ equality, one can
         define a projection operator $\tilde{P}_{\leq_{A,R}}.$
         Here the definition is slightly more complex as the signs
         of the two components to be compared must be taken into
         account.  Similar relations hold for the
         imaginary components of the rational string states.

          The definitions of $=_{A}$ and $\leq_{A}$ can also be applied
         to states $\psi,\psi^{\p}$ that are linear superpositions
         of rational string states. The probability, $P_{\psi =_{A}\psi^{\p}},$
         that $\psi=_{A}\psi^{\p}$ is given by \begin{equation}\label{probpsi=A}
         \begin{array}{l}P_{\psi =_{A}\psi^{\p}}=\sum_{\asbt}
         \sum_{\asbtp}\{ \\ \hspace{1cm}|\langle \asbt|\psi\rangle|^{2}
         |\langle\asbtp|\psi^{\p}\rangle|^{2}\\ \hspace{0.5cm}\times
         \langle\asbt|\tilde{P}_{=_{A}}|\asbtp\rangle\}\end{array}\end{equation}
         where\begin{equation}\label{PRI} \tilde{P}_{=_{A}}=
         \tilde{P}_{=_{A,R}}\tilde{P}_{=_{A,I}}\end{equation} is a
         product of projection operators for the real and imaginary
         components. The validity of this expression also depends on
         the fact that $\tilde{P}_{=_{A,R}}$ and $\tilde{P}_{=_{A,I}}$ commute.
         Expressions for the probability that $\psi\leq_{A}\psi'$ will
         not be given here as they are similar.

        \subsection{Arithmetic Operations}\label{AO}

         The basic arithmetic operations to be described are
         addition, subtraction, multiplication, and division
         to arbitrary accuracy. As is well known states of
         finite qubit strings are not closed under division.
         However, one can implement division to any finite
         accuracy on these states. Let $\tilde{O}_{A}$ be a unitary
         operator for describing these arithmetic operations.
         One has\begin{equation}\label{defO}\begin{array}{l}
         \tilde{O}_{A}|\asbt\rangle |\asbtp\rangle \\
         =|\asbt\rangle|\asbtp\rangle|\asbtpp\rangle
         \end{array}\end{equation} where\begin{equation}\label{stOAstp}
         |\asbtpp\rangle =_{A}|(\asbt)O_{A}(\asbtp)\rangle.\end{equation}
        The expression $|(\asbt)O_{A}(\asbtp)\rangle$ is a useful notation that
        uses $O$ inside $|-,-\rangle$ to represent the rational
        string state resulting from carrying out the operation
        $O_{A}$. To ensure unitarity for $\tilde{O}_{A}$ the first
        two states are repeated with the resultant state created.
        Also the lattice location $(m'',h'')$ of this state differs
        from those of the first two states.

        Application of the arithmetic operations to states that are linear
         superpositions of the string states creates entangled
         states. One has \begin{equation}\label{Oentngl}
        \begin{array}{l}\tilde{O}_{A}\psi\psi^{\p}
         = \sum_{\a\g,s,\gp,t}\sum_{\g_{1}, \sp,\gp_{1},\tp}
        \langle\asbt|\psi\rangle\langle\asbtp|\psi^{\p}\rangle \\
        \hspace{1cm}|\asbt\rangle|\asbtp\rangle|(\asbt)O_{A}(\asbtp)\rangle.
        \end{array}\end{equation} Taking the trace over the $\psi$
        and $\psi^{\p}$ components gives a mixed state
        \begin{equation}\label{rhoplus}\begin{array}{l}
        \rho_{\psi O_{A}\psi^{\p}}=\sum_{\g,s,\gp,t}\sum_{\g_{1}, \sp,\gp_{1},\tp}
        |\langle\asbt|\psi\rangle|^{2}\\ \hspace{.5cm}\times
        |\langle\asbtp|\psi^{\p}\rangle|^{2}\rho_{(\asbt)O_{A}(\asbtp)}
        \end{array}\end{equation} to represent the result of the operation.

        The definitions of $\tilde{O}_{A}=\tilde{+}_{A},\tilde{-}_{A},
        \tilde{\times}_{A},\tilde{\div}_{A,\ell}$ are all based on the
        use of successor operators, one for each $j$.  To save on
        notation, $x^{\dag}_{i,j,h},x_{i,j,h}$ will denote
        $\ad_{i,j,h},a_{i,j,h}$ or $\bd_{i,j,h},b_{i,j,h}$ for the
        real $X=R$ or imaginary $X=I$ components of the state.  The
        successor operator for each $X,j,h$ is defined by an
        iterative expression
        \begin{equation}\label{DefNXjh}\begin{array}{l}
        \tilde{N}_{X,j,h}=\xd_{1,j,h}x_{0,j,h}+\xd_{1,j+1,h}
        \xd_{0,j,h}x_{1,j,h}P_{X,unocc,j+1,h}\\ +P_{X,occ,j+2,h}
        P_{X,occ,j+1,j}\tilde{N}_{X,j+1,h}\xd_{0,j,h}x_{1,j,h}.
        \end{array}\end{equation}  Here $P_{X,occ,j+1,h},\;
        P_{X,unocc,j+1,h}$ are projection operators for finding
        site $j+1,h$ occupied or unoccupied by a type $x$ qubit. The
        adjoint $\tilde{N}^{\dag}_{X,j,h}$ is defined by
        \begin{equation}\label{DefNdagXjh}\begin{array}{l}
        \tilde{N}^{\dag}_{X,j,h}=\xd_{0,j,h}x_{1,j,h}+P_{X,unocc,j+1,h}
        \xd_{1,j,h}x_{0,j,h}x_{1,j+1,h}\\ +\xd_{1,j,h}x_{0,j,h}
        \tilde{N}^{\dag}_{X,j+1,h}P_{X,occ,j+1,h}
        P_{X,occ,j+2,j}.\end{array}\end{equation}

        The action of $\tilde{N}_{X,j,h}$ on a state
        $|\g,(m,h),s;\gp,t\rangle$ in which site
        $j,h$ is occupied, creates a new state that corresponds to
        arithmetic addition of $2^{j-m}$ for $X=R$ and $i2^{j-m}$
        for $X=I.$ The action of the adjoint $\tilde{N}^{\dag}_{X,j,h}$ on the
        same state corresponds to subtraction of $2^{j-m}$ for $X=R$
        and $i2^{j-m}$ for $X=I.$ It is useful to note, too, that
        \begin{equation}\label{Nsqrd}(\tilde{N}_{X,j,h})^{2}
        |\asbt\rangle=\tilde{N}_{X,j+1,h}|\asbt\rangle\end{equation}
        if sites $j,h$ and $j+1,h$ are occupied by $x$ qubits.
        This corresponds to the observation that
        $2^{j-m}+2^{j-m}=2^{j+1-m}.$

        One also has to extend the definition of $\tilde{N}_{X,j,h}$
        to include cases where site $j,h$ is unoccupied. Examples at
        either end of a string are $110.1+ 0.000001$ or $110.1 +
        100000.0$  To accommodate this one defines an operator
        $Z_{X,m,j,h}$ by \begin{equation}\label{Zdef}
        Z_{X,m,j,h} = P_{X,occ,(j,h)}+\left(
        \begin{array}{l}Z_{X,m,j-1,h}\xd_{0,j,h}P_{X,unocc,(j,h)}
        \mbox{ if $j>m$} \\ Z_{X,m,j+1,h}\xd_{0,j,h}P_{X,unocc,(j,h)}
        \mbox{ if $j<m$} \\ \xd_{0,m,h}P_{X,unocc,(m,h)}\mbox{ if
        $j=m$.}\end{array}\right. \end{equation}

        Note that, as defined, $Z_{X,m,j,h}$ is a many-one operator as
        it creates the same rational string states from states with
        different numbers of terminal $0s$.  To avoid this source of
        irreversibility one needs to first copy the state on which
        $Z_{X,m,j,h}$ will act. To this end a a copying operator
        $C$ is defined where\begin{equation}\label{copy} C|\g,(m,h),s;
        \gp,t\rangle=|\g,(m,h),s;\gp,t\rangle|\g,(m,h'),s;\gp,t\rangle.
        \end{equation} The state $|\g,(m,h'),s;\gp,t\rangle$ is a
        copy of $|\g,(m,h),s;\gp,t\rangle$ that is located along $h'$
        instead of $h$. One should note that, because of the no-cloning
        theorem \cite{Wootters1}, $C$ does not copy a state that is a
        linear superposition of the basis states $|\asbt\rangle.$
        Instead it creates an entangled superposition of pairs of
        basis states where each pair differs only in the value of
        the parameter $h$.

        The action of $Z_{X,m,j,h'}$ on a pair of
        states $|\asbt\rangle|\asbtp\rangle$ does nothing if site
        $j,h'$ is occupied by a type $x$ qubit. otherwise it adds a
        string of $x$ qubits  along $h'$ in state $|0\rangle$ that
        terminates at site $j,h'.$ The addition is from $u$ or $u'$
        to $j$ if $j>m,$ from $l$ or $l'$ to $j$ if $j<m,$ and just at
        $m,h'$ if $j=m.$

        The combination $\tilde{N}_{X,j,h}Z_{X,m,j,h}$ is quite useful.
        This follows from the observation that
        the state $\cd_{\g,(m,h)},(\ad)^{\bar{0}_{[(j,h),(m,h)]}}
        \ad_{1,j+1,h}|0\rangle,$ corresponding to the number $2^{j-m},$
        can be expressed as $\tilde{N}_{R,j,h}Z_{R,m,j,h}
        \cd_{\g,(m,h)}|0\rangle.$ It can also be expressed
        as\begin{equation}\label{stringNZ} \begin{array}{l}
        \cd_{\g,(m,h)},(\ad)^{\bar{0}_{[(j,h),(m,h)]}}
        \ad_{1,j-1,h}|0\rangle =\cd_{\g,(m,h)}(\tilde{N}_{R,u,h}
        Z_{R,m,u,h})^{s(u)}\\\times(\tilde{N}_{R,u-1,h}Z_{R,m,u-1,h})^{s(u-1)}
        \cdots(\tilde{N}_{R,l,h}Z_{R,m,l,h})^{s(l)}
        |0\rangle\end{array}\end{equation} Here $[(l,h),(u,h)]$ is the
        domain of $s$ with $l=j+1$, $u=m,$ and $s(k,h)=0$ for all
        $m\geq  j\geq l+1$ and $s(l,h)=1.$

        This can be extended to any state $|\g,s\rangle$ to give
        \begin{equation}\label{NZs}\begin{array}{l}|\g,s\rangle=\cd_{\g,m,h}
        \ads|0\rangle=\cd_{\g,m,h}(\tilde{NZ}_{R,[l,u],h})^{s}
        |0\rangle \\=\cd_{\g,m,h}(\tilde{N}_{R,u,h}Z_{R,m,u,h})^{s(u,h)}\cdots
        (\tilde{N}_{R,l,h}Z_{R,m,l,h})^{s(l,h)}|0\rangle.
        \end{array}\end{equation}  One has a similar expression for the
        imaginary component:\begin{equation}\label{NZt}\begin{array}
        {l}|\g',t\rangle=\dd_{\g',m,h}\bdt|0\rangle=\dd_{\g',m,h}
        (\tilde{NZ}_{I,[l',u'],h})^{s}|0\rangle\\=\dd_{\g',m,h}
        (\tilde{N}_{I,u',h}Z_{I,m,u',h})^{t(u',h)}\cdots
        (\tilde{N}_{I,l',h}Z_{I,m,l',h})^{t(l',h)}|0\rangle.
        \end{array}\end{equation}

        This can all be put together to define arithmetic addition
        and subtraction. One has for $\tilde{O}_{A}$ where $O=+$ or
        $O=-$,\begin{equation}\label{DefOA}\begin{array}{l}
        \tilde{O}_{A}|\asbt\rangle|\asbtp\rangle=\tilde{O}'_{A}
        |\asbt\rangle C|\asbtp\rangle \\ =|\asbtp\rangle\tilde{O}'_{A}
        |\asbt\rangle|\asbtp\rangle \\ =|\asbtp\rangle|\asbt\rangle|(\asbt)O'_{A}
        (\asbtp)\rangle.\end{array}\end{equation} Here the state
        $|\asbtp\rangle$ is copied and $\tilde{O}'_{A}$ acts on
        $|\asbt\rangle$ and the produced copy state.

        To define $|(\asbt)O'_{A}(\asbtp)\rangle$ it is easier to
        consider just the real component as the treatment for the
        imaginary component is the same for addition and subtraction.
        One also needs to consider separately the different signs of
        the string states.

        For $\g =\g_{1} =+$ or $\g =\g_{1} =-$ and $O=+$,
        \begin{equation}\label{++plus} |(\g,(m,h),s)+_{A}(\g_{1},(m,h")s')\rangle=
        \cd_{\g_{1},(m,h'')}(\tilde{NZ}_{R,[l,u],h''})^{s}\adsp|0\rangle.
        \end{equation}where, following Eq. \ref{NZs},
        \begin{equation}\label{NZs1}(\tilde{NZ}_{R,[l,u],h''})^{s}=
        (\tilde{N}_{R,u,h''}Z_{R,m,u,h''})^{s(u,h)}\cdots
        (\tilde{N}_{R,l,h''}Z_{R,m,l,h''})^{s(l,h)}.\end{equation}
        Note that the powers $s(u,h)\cdots s(l,h)$ are obtained from the qubit
        string along $h$ but the $NZ$ factors are applied to the
        copy string along $h''.$

        For $\g=-$ and $\g_{1}=+,$ there are two cases to consider. If
        $W_{R}|\g,(m,h),s\rangle\leq_{A}|\g_{1},(m,h'),s'\rangle$ where
        \begin{equation}\label{defW}W_{R}=\sum_{m,h}\cd_{-,m,h}c_{+,m,h}+
        \cd_{+,m,h}c_{-,m,h}\end{equation} is a sign change operator
        for the real component, then\begin{equation}\label{-+plus}
        |(\g,(m,h),s)+_{A}(\g_{1},(m,h'')s')\rangle=\cd_{\g_{1},(m,h'')}
        (\tilde{N^{\dag}Z}_{R,[l,u],h''})^{s}\adsp|0\rangle\end{equation}
         and\begin{equation}\label{NdagZs}
        (\tilde{N^{\dag}Z}_{R,[l,u],h''})^{s}=
        (\tilde{N}^{\dag}_{R,u,h''}Z_{R,m,u,h''})^{s(u,h)}\cdots
        (\tilde{N}^{\dag}_{R,l,h''}Z_{R,m,l,h''})^{s(l,h)}.\end{equation}
        This uses the observation that addition of a negative number
        is equivalent to subtraction of a positive one.

        If $W_{R}|\g,(m,h),s\rangle\geq_{A}|\g_{1},(m,h'),s'\rangle$
        then\begin{equation}\label{-+plus1}\begin{array}{l} |(\g,(m,h),s)+_{A}(\g_{1},
        (m,h'')s')\rangle=W_{R}|(\g_{1},(m,h''),s')+_{A}(\g,(m,h),s\rangle \\
        \hspace{1cm}=W_{R}\cd_{\g,(m,h)}(\tilde{N^{\dag}Z}_{R,[l,u],h})^{s'}\ads|0\rangle.
        \end{array}\end{equation} This expresses the fact that $A+(-B)= -(B+(-A)).$

        This covers all the cases for addition because the case for
        $\g=+$ and $\gp=-$ is obtained from the results above. Also
        all the cases for subtraction are included because
        $A-B=A+(-B).$  This is shown in the above where
        $\tilde{N}^{\dag}_{R,j,h}$, which corresponds to subtraction
        of $2^{j-m}$, is used.

        These results also extend to the imaginary part in an obvious way.
        $(\tilde{NZ}_{R,[l,u],h''})^{s}$ becomes \begin{equation}
        \label{NZthpp}(\tilde{NZ}_{I,[l',u'],h''})^{t}=
        (\tilde{N}_{I,u',h''}Z_{I,m,u',h''})^{t(u',h)}\cdots
        (\tilde{N}_{I,l',h''}Z_{I,m,l',h''})^{t(l',h)}\end{equation}
        and $W_{R}$ is replaced by\begin{equation}\label{WIdef}
        W_{I}=\sum_{m,h}(\dd_{+,m,h}d_{-,m,h}+\dd_{-,m,h}d_{+,m,h}).
        \end{equation}  Addition or subtraction of both the real and
        imaginary components is defined using the above definitions
        for all the cases that arise.

        The definitions given have the axiomatic property that $0$
        is an additive identity.  To see this one notes from Eq.
        \ref{++plus} that if $s(j)=0$ for all $j$ in $[l,u]$ then
        \begin{equation}\label{0addid}\cd_{\g_{1},m,h''}
        (\tilde{N}_{R,[l,u],h''}Z_{R,m,[l,u],h''})^{s}
        \adsp|0\rangle=\cd_{\g_{1},m,h''}\adsp|0\rangle.\end{equation}
        If $s'(j)=0$ for all $j$ in $[l',u']$ then \begin{equation}
        \label{0addid1}\cd_{\g_{1},m,h''}(\tilde{N}_{R,[l,u],h''}Z_{R,m,[l,u],h''})^{s}
        \adsp|0\rangle=\cd_{\gp,m,h''}\ads|0\rangle.\end{equation} This follows
        directly from Eq. \ref{NZs}.

        Multiplication is more complex. Results given in more
        detail in \cite{BenRCRNQM} are summarized here. Based
        on Eq. \ref{defO}, the state resulting from multiplication
        can be represented as $|(\asbt)\times_{A}(\asbtp)\rangle.$
        Following the usual rules of multiplication of complex numbers
        gives\begin{equation}\label{timesN} \begin{array}{l}|(\asbt)
        \times_{A}\asbtp\rangle =|[(\g,s\times_{A}(\g_{1},s'
         ))+_{A}((\gp,t)\times_{A}(\gp_{1},t'))] \\;
        +_{A}[((\g,s)\times_{A}(\gp_{1},t'))+_{A}((\gp,t)
        \times_{A}(\g_{1},\sp)]\rangle.\end{array}\end{equation}
        This definition shows multiplication for complex string states
        defined by four component multiplications. Then the resulting
        real components are added together as are the imaginary components.

        The four component multiplications are defined by
        \begin{equation}\label{compmult}
        \begin{array}{l}|(\g,s)\times_{A}(\gp,s')\rangle=\cd_{\g}\ads
        \times_{A}\cd_{\gp}\adsp|0\rangle
        =\cd_{\g''}\adspp|0\rangle\\ \hspace{1cm}\mbox{ $\g''=+$ if
        $\g=\gp,$ $\g'' =-$ if $\g\neq\gp$}\\
        |(\gp,t)\times_{A}(\gp_{1},t')\rangle =\dd_{\gp}\bdt
        \times\dd_{\gp_{1}}\bdtp|0\rangle
        =|\cd_{\g''}(\ad)^{s_{1}}|0\rangle\\ \hspace{1cm}\mbox{ $\g''=+$ if
        $\g\neq\gp,$ $\g'' =-$ if $\g=\gp$}\\ |(\gp,t)\times_{A}(\g,s)
        \rangle=\dd_{\gp}\bdt\times_{A}\cd_{\g}\ads|0\rangle
        =|\dd_{\g''}\bdtpp|0\rangle \\ \hspace{1cm}\mbox{ $\g''=+$ if
        $\g=\gp,$ $\g'' =-$ if $\g\neq\gp$}\\ |\cd_{\g}\ads
        \times_{A}\dd_{\gp}\bdt|0\rangle
        =|\dd_{\g''}\bdtpp|0\rangle \\ \hspace{1cm}\mbox{ $\g''=+$ if
        $\g=\gp,$ $\gp =-$ if $\g\neq\gp$}.
        \end{array}\end{equation}Location labels have been left of
        off $\cd$ and $\dd$ to save on space.

        There are four multiplications to consider.  However it is
        sufficient to examine only one as the others follow the same
        rules. A unitary shift operator $T$ is useful here where $T$
        satisfies the commutation rule\begin{equation}\label{Tdef}\begin{array}{c}
        T\ad_{i,j,h}=\ad_{i,j+1,h}T \\Ta_{i,j,h}=a_{i,j+1,h}T
        \\T\bd_{i,j,h}=\bd_{i,j+1,h}T,
       \\Tb_{i,j,h}=b_{i,j+1,h}T\\T|0\rangle=|0\rangle.
       \end{array}\end{equation}  These equations show that $T$ increases
       the value of $j$ to $j+1$.  Conversely taking the adjoints of these
       equations shows that $T^{\dag}=T^{-1}$ decreases the value of
       $j$ to $j-1.$ The use of $T$ and $T^{-1}$ derive from the
       observation that their actions correspond to multiplying
       and dividing by $2.$ The state
        $\adspp|0\rangle$ is defined from $\ads|0\rangle$ and $\adsp|0\rangle$ by
        \begin{equation}\label{multdef}\begin{array}{l}\adspp|0\rangle=(T^{u-m}\adsp)^{s(u)}
        +_{A}((T^{u-1-m}\adsp)^{s(u-1)}\\ +_{A}(\cdots +_{A}(T^{l-m}\adsp))^{s(l)}
        )|0\rangle.\end{array}\end{equation}Here\begin{equation}\label{Tkasp}
        T^{k}\adsp=(\ad)^{s'_{k}}T^{k}\end{equation}where $s'_{k}(j+k)=s'(j)$ for
        $l\geq j\geq u.$

        This can be expressed using the successor operators $\tilde{N}_{R,j,h}
        Z_{R,m,j,h}$ defined by Eqs. \ref{++plus}-\ref{-+plus} and
        \ref{-+plus1}.  From Eq. \ref{NZs} one can substitute
        $(\tilde{NZ}_{R,[l,u],h})^{s'}$ for $\adsp$ and use
        the definition of addition,Eq. \ref{++plus}, to
        obtain\begin{equation}\label{adsppNZ}\begin{array}{l}
        \adspp|0\rangle=(T^{u-m}(\tilde{NZ}_{R,[l',u'],h})^{s'})^{s(u)}
        ((T^{u-1-m}(\tilde{NZ}_{R,[l',u'],h})^{s'})^{s(u-1)}\\ (\cdots (T^{l-m}
        (\tilde{NZ}_{R,[l',u'],h})^{s'}))^{s(l)}
        )|0\rangle.\end{array}\end{equation}

        This equation looks complex because it expresses the steps
        one goes through, using repeated additions to carry out
        multiplication.  It is a quantum theoretical expression of
        the usual multiplication rule shown by the following simple
        example.  Let $s$ and $s'$ be $0,1$ functions with
        respective integer interval domains $[l,u],[l',u']$ and with
        the "binal" point at $m$ where $m$ is in both domains.  Then
        $s\times s' =\sum_{j=l}^{u}s(j)(2^{j-m}\times s').$ Note
        that $s(j)$ appears as a factor here instead of an exponent.

        Multiplication for the other three entries in Eq.
        \ref{compmult} follows the same rules.  Conversion of type
        $b $ qubits to type $a$ qubits and conversely can be
        expressed explicitly by the use of type change operators.
        This will not be done here as it adds nothing new.

         As is the case for rational string numbers, the
         string states described here are not closed under
        division. However they are closed under division to any
        accuracy $|+,-\ell\rangle=\ad_{+,-\ell}|0\rangle.$ For any
        pair of  states $|\asbt\rangle\neq
        |0\rangle$ and $|\asbtp\rangle,$ the state
        $|(\asbtp/\asbt)_{\ell}\rangle$ is defined by
        \begin{equation}\label{divdef}|(\asbtp/\asbt)_{\ell}
        \rangle=|(+,1/\asbt)_{\ell}\times_{A}(\asbtp)\rangle\end{equation}
        where the $\ell$ inverse  state
        $|(+,1/\asbt)_{\ell}\rangle$ is defined by
       \begin{equation}\label{ellinvcplx}\begin{array}{l}
        |(+,1/\asbt)_{\ell}\rangle \\ \hspace{.2cm}=|(+,1/[(\g s\times_{A}\g
        s)+(\gp t\times_{A} \g^{\p\p}t)])_{\ell}\times_{A} (\g s,\g^{\p\p}t)\rangle.
        \end{array}\end{equation} Here $\g^{\p\p}\neq\gp$ and

        From Eq. \ref{compmult} one sees that the denominator
        state $|(+,1/[(\g s\times_{A}\g
        s)+(\gp t\times_{A} \\g^{\p\p}t)])_{\ell}\rangle$ of
        the $\ell$ inverse is a positive, real state. Thus to
        complete the definition of $\ell$ division it suffices to
        define the state that is the $\ell$ inverse of a
        positive, real  state $|+, s\rangle\neq |+,0\rangle.$  To
        this end $|(+1/+, s)_{\ell}\rangle=(+,\adsp|0\rangle$ is
        defined by two
        conditions:\begin{equation}\label{definv}\begin{array}{l}
       |(+,1)-_{A}(+,\bar{0}_{[m,m-\ell+1]}1_{m-\ell})\rangle \\
       \hspace{1cm}\leq_{A}|(+1/+, s)_{\ell}\times_{A}(+,
       s)\rangle\leq_{A}|+,1\rangle,\end{array}\end{equation} and, if
       $|+,s^{\p\p}\rangle$ is another state such that $|(+1/+,
       s^{\p\p})_{\ell}\times (+, s^{\p\p})\rangle$ satisfies the above double
       inequality, then the smallest $j$ value in $\sp$ where
       $\sp(j)\neq 0$ is larger than that in $s^{\p\p}.$

       The first condition states that $(1/s)_{\ell}\times s$ must lie between
       $1-2^{-\ell}$ and $1$.  The second condition states that
       $(1/s)_{\ell}$ is the largest number to satisfy the first
       condition.

        As an example, assume $m=0$ and let $|+, s\rangle=
        |+,\ad_{1,2}\ad_{0,1}\ad_{1,0}|0\rangle$ and $\ell =7.$ Among many
        others, the states $|+,(\ad)^{s_{1}}|0\rangle$ and
        $|+,(\ad)^{s'_{1}}|0\rangle$ where $$\begin{array}{c}
        s_{1}(j)=1 \mbox{ if }j=-3,-4,-7,-9,-10,-11
        \mbox{ and $s_{1}(j)=0$ for all other $j$ in $[-11,0],$}\\
        s'_{1}(j)=1 \mbox{ if }j=-3,-4,-7,-8
        \mbox{ and $s'_{1}(j)=0$ for all other $j$ in
        $[-8,0],$}\end{array}$$ both satisfy Eq. \ref{definv}
       for $\ell = 7.$ Also a more restricted s'' with $s''(j) =1$
       if $j= -3,-4,-7$ does not satisfy the conditions.
       Based on this, the unique $\ell$
       inverse state for $\ell=7$ is $|+,(\ad)^{s'_{1}}|0\rangle$ as
       $-8$ is greater than $-11.$

       In binary numbers the example says that
       $0.00110010111$ and $0.00110011$ are among the many results accurate
       to $0.0000001=2^{-7}$ of the division $1/101.0$ to
       an accuracy of $2^{-7}.$  The second condition says
       keep the largest one, which is $0.00110011.$
       Additional details on the explicit construction of the
       state $|(+,1/+, s)_{\ell}\rangle$ are given in \cite{BenRCRNQM}.

        So far,  states of the form $|\g,s;\gp,t\rangle$ and their superpositions
        have been defined,, along with the arithmetic
        relations $=_{A},\leq_{A}$ and the operations $\tilde{+}_{A},
        \tilde{-}_{A},\tilde{\times}_{A},\tilde{\div}_{A,\ell}.$
        The arithmetic relations and operations are defined on these
        states in order to show that these states do represent
        complex rational numbers. This follows from showing that the
        relations and operations satisfy the requisite axioms for
        complex natural numbers.  Included are the commutativity and associativity
        of $\tilde{+}_{A},\tilde{-}_{A},\tilde{\times}_{A},$ the identity property
        of $|0\rangle$ and $|+,1\rangle=\cd_{+,m}\ad_{1,m}|0\rangle$
        for $\tilde{+}_{A}$ and$\tilde{\times}_{A},$ the
        distributivity of $\tilde{\times}_{A}$ over
        $\tilde{+}_{A},$ etc. These are discussed in \cite{BenRCRNQM}.

          \section{Cauchy Sequences of States in $\mathcal{F}$} \label{CSSF}

          The rest of this paper is concerned with sequences of
          states  that satisfy the Cauchy condition.  Sequences of
          states are defined to be functions  $\psi:\{1,2,\dots\}
          \rightarrow \mathcal{F}$ where $\psi_{n}$ is a state in
          $\mathcal{F}.$ If the states $\psi_{n}$ are basis states
          $|\g_{n},s_{n};\gp_{n},t_{n}\rangle$ the sequence will be
          denoted by $\{|\g_{n},s_{n};\gp_{n},t_{n}\rangle\}$
          instead of the more general form $\{\psi_{n}\}.$ In the
          general case $\{\psi_{n}\}$ is a sequence of normalized
          states where \begin{equation}\label{psinexp}
          \psi_{n}=\sum_{\g,,s,\gp,t}|\g,(m,h),s,
          \gp,t)\rangle\langle\g,(m,h),s,\gp,t)|\psi_{n}\rangle
          \end{equation}  and $\langle\psi_{n}|\psi_{n}\rangle=1$
          for each $n$. Note that the sum is over states of qubit
          strings of all finite lengths.

          One condition implied by the definition of Eq.
          \ref{psinexp} is that the value of $(m,h)$ is the
          same for each state in the sequence. This is also the case
          for the sequence $\{|\g_{n},s_{n};\gp_{n},t_{n}\rangle\}$
          of basis states. This is a locality condition that places
          the sign qubits at the same location in $I\times I$ for
          each $n.$ One could relax this condition by including a
          sum over $(m,h)$ in Eq. \ref{psinexp}, but this will not
          be done here.

          Sequences that satisfy the Cauchy conditions are of interest
          here as the goal (Section \ref{RRCNQT}) is to show that
          (equivalence classes of) these Cauchy sequences are
          real or complex numbers. For the real numbers one must show
          that the equivalence classes are a complete, ordered field. For
          the complex numbers the required properties are those
          of a complete, algebraically closed, field \cite{Hewitt,Randolph}.

          The reason that one studies Cauchy sequences of states
          instead of convergent sequences is that the definition of the Cauchy
          condition is based directly on properties of states in
          $\mathcal F.$  Convergence is not used  because the sequences themselves
          are not elements of $\mathcal F.$  Thus convergence to a
          sequence has no meaning in $\mathcal F.$

          \subsection{The Cauchy Condition}\label{TCC}

          A Cauchy sequence of rational numbers $\{x_{n}:n=1,2,\cdots\}$
          is a sequence that satisfies the Cauchy condition:
          \cite{Hewitt,Randolph} \begin{equation}\label{CCNos}\begin{array}{c}
          \mbox{For each $\ell$ there is an $h$ such that for all $m,n>h$}
          \\ |x_{n}-x_{m}|<2^{-\ell}.\end{array}\end{equation}

         This definition can also be applied to sequences
         $\{x_{n}=u_{n}+iv_{n}:n=1,2\cdots\}$  of complex rational numbers.
         In this case  $|x_{n}-x_{m}|<2^{-\ell}$ is expressed here
          as two separate conditions, $|u_{n}-u_{m}|<2^{-\ell}$
          and $|v_{n}-v_{m}|<2^{-\ell},$ for the real and imaginary
          parts of the sequence numbers. It is possible to combine
          the two conditions into one, but this will not be done
          here.

           The Cauchy condition can be applied to basis states of qubit strings.
           Let $\{|\g_{n},s_{n};\gp_{n},t_{n}\rangle\}$ be a
           sequence of states of qubit strings. The Cauchy condition
          for these states is \begin{equation}\label{cauchyri}
          \begin{array}{l}\hspace{0.5cm}\mbox{ For each
          $\ell$ there is an $h$ where for all $j,k>h$} \\
          \hspace{1cm} |(|\g_{j}s_{j}-_{A}\g_{k}s_{k}|_{A})\rangle
          <_{A}|+,-\ell\rangle\mbox{ and }\\ \hspace{1cm}
          |(|\gp_{j}t_{j}-_{A}\gp_{k}t_{k}|_{A})\rangle<_{A}|+,-\ell\rangle
          \end{array} \end{equation} Here two separate conditions are
          used, one for the real component and one for the imaginary component.

          Here $|(|\g_{j}s_{j}-_{A}\g_{k}s_{k}|_{A})\rangle$ is the
          state that is the arithmetic absolute value of the state
          resulting from the arithmetic subtraction of
          $|\g_{k},s_{k}\rangle$ from $|\g_{j}s_{j}\rangle.$ The
          subscripts $A$ are used to indicate that the operations
          are arithmetic and not the usual quantum theory operations.

          The absolute value state $|(|(\g_{k},s_{k})-_{A}
          (\g_{j},s_{j})|_{A})\rangle$ is represented in terms of
          AC operators by (Eq. \ref{-+plus})\begin{equation}\label{absdiff}
          \begin{array}{l}\cd_{+,(m,h'')}(\tilde{N^{\dag}Z}_{R,[l_{k},u_{k}],h''})^{s_{k}}
          (\ad)^{s_{j}}|0\rangle\mbox{ if $|+,s_{j}\rangle \geq_{A}|+,s_{k}
          \rangle$ }\\ \cd_{+,(m,h'')}(\tilde{N^{\dag}Z}_{R,[l_{j},u_{j}],h''})^{s_{j}}
          (\ad)^{s_{k}}|0\rangle \mbox{ if $|+,s_{k}\rangle
          \geq_{A}|+,s_{j}\rangle$} \end{array}\end{equation} if
          $\g_{k}=\g_{j}.$  If $\g_{k}\neq \g_{j},$ $|(|(\g_{k},s_{k})-_{A}
          (\g_{j},s_{j})|_{A})\rangle$ is represented by\begin{equation}
          \label{gkn=gj}\cd_{+,(m,h'')}(\tilde{NZ}_{R,[l_{k},u_{k}],h''})^{s_{k}}
          (\ad)^{s_{j}}|0\rangle.\end{equation} Similar relations with
          $t$ replacing $s$ hold for the imaginary component. In this
          case one is using $|\pm ix-\pm iy|=|i(\pm x-\pm y)|=|\pm x-\pm y|.$

          Note that this definition of the Cauchy condition is local
          in the sense that it is defined for sequences at three locations
          $(m,h),(m,h')$ and $(m,h'').$ Each state in the sequence
          is a state of a qubit string at $(m,h)$ (Eq. \ref{psinexp}).
          Since $A$ subtraction is a binary operation, the
          sequence must be copied to another $h$ location, $(m,h').$
          The result of the subtraction is a third sequence at
          another $h$ location, $(m,h'').$ Converting to the
          absolute  value is simply a change in the state of the
          sign qubit(s) and does not create a new sequence.
          Finally, the state $|,-\ell\rangle$ is at another location
          $(m,h''').$

          A global definition of the Cauchy condition can be given by summing
          over $(m,h),(m',h'),(m'',h''),(m''',h''')$ with the restriction
          that the $h$ parameters are pairwise distinct. However this
          will not be done here. In the following these location
          labels will be suppressed.

          The Cauchy condition can also be defined for sequences
          $\{\psi_{n}\}$ where $\psi_{n}$ is a normalized
          state given by Eq. \ref{psinexp}. These definitions make use of
          $R$ and $C$ as they are based on probabilities obtained
          from the expansion coefficients of the states.  The
          coefficients are elements of $C$.

          A sequence  $\{\psi_{n}\}$ is defined  to be a
          Cauchy sequence if the probability is unity that the
          Cauchy condition is true for the sequence. This probability
          is obtained by applying the conditions of Eqs. \ref{cauchyri}
           to the components of $\psi_{j},\psi_{k}$
          and summing over all components that satisfy the conditions.
          To see this for Eq. \ref{cauchyri}, let
          $\tilde{|Re|},\tilde{|Im|}$ be operators acting on
          the real and the imaginary parts of the states that are
          defined by\begin{equation}\label{Re-IM} \begin{array}{l}
          \tilde{|Re|}\tilde{|Im|}|\g,s,\g_{1},t\rangle |\gp,\sp,\gp_{1},
          \tp\rangle\\=|\g,s,\g_{1},t\rangle|\gp,\sp,\gp_{1},\tp\rangle
          |(|\g,s-_{A}\gp,\sp|_{A})\rangle|(|\g_{1},t-_{A}\gp_{1},t'|_{A})
          \rangle.\end{array}\end{equation} Here lattice locations of the
          qubit strings are suppressed to avoid notation clutter.

          Let $\tilde{P}_{Re\leq \ell},\tilde{P}_{Im\leq \ell}$ be
          projection operators for positive states being
          arithmetically less than $|+,-\ell\rangle,$
          \begin{equation}\label{PReIm}\begin{array}{l}\tilde{P}_{Re\leq
          \ell}|+,s\rangle =\left\{\begin{array}{l}|+,s\rangle
          \mbox{ if $|+,s\rangle\leq_{A}|+,-\ell\rangle$}\\
          \mbox{$0$ if $|+,s\rangle >_{A}|+,-\ell\rangle$}\end{array}\right.\\
          \tilde{P}_{Im\leq\ell}|+,t\rangle =\left\{\begin{array}{l}|+,t\rangle
          \mbox{ if $|(|+,t|_{A})\rangle\leq_{A}|+,-\ell\rangle$}\\
          \mbox{$0$ if $|(|+,t|_{A})\rangle >_{A}|+,-\ell\rangle.$}\end{array}
          \right.\end{array}\end{equation} Here the state $|(|+,t|_{A})
          \rangle=\cd_{+}(\ad)^{t}|0\rangle$ is a real rational state, so
          it can be directly compared with $|+,-\ell\rangle.$

          Putting these results together and letting\begin{equation}\label{theta}
          \Theta_{\psi_{j}\psi_{k}}=\tilde{|Im|}\tilde{|Re|}\psi_{j}\psi_{k}
          \end{equation} gives\begin{equation}\label{probjkl}\begin{array}{l}
          P^{\{\psi_{n}\}}_{j,k,\ell}=\langle\Theta_{\psi_{j}\psi_{k}}|
          \tilde{P}_{Re<\ell}\tilde{P}_{Im<\ell}|\Theta_{\psi_{j}\psi_{k}}\rangle
          \\ =\sum_{\g,s,\gp,t}\sum_{\g_{1},\sp,\g'_{1},t'}|\langle\g,s,\gp,t
          |\psi_{j}\rangle|^{2}|\langle\g_{1},\sp,\g'_{1},t'|\psi_{k}\rangle|^{2}: \\
          |(|\g,s-_{A}\g_{1},s'|_{A})\rangle\leq_{A}|+,-\ell\rangle \mbox{ and}\\
          |(|\gp, t-_{A}\gp_{1},\tp|_{A})\rangle\leq_{A}|+,-\ell\rangle.\end{array}
          \end{equation} The projection operator product
          $\tilde{P}_{Re<\ell}\tilde{P}_{Im<\ell}$ limits the
          sums to those component states that satisfy the Cauchy
          conditions for both the real and imaginary components.
          $P^{\{\psi_{n}\}}_{j,k,\ell}$ is the probability that $\{\psi_{n}\}$
          satisfies these conditions at $j$ and $k$, i.e.
          for $\psi_{j}$ and $\psi_{k}.$

          One now has to account for the quantifiers in the
          definition of the Cauchy condition. This is done in steps.
          The probability, $P^{\{\psi_{n}\}}_{h,\ell},$ that $\{\psi_{n}\}$
          satisfies the conditions for all $j,k>h$ is given by
         \begin{equation}\label{probellh}P^{\{\psi_{n}\}}_{h,\ell}=
         \liminf_{j,k>h}P^{\{\psi_{n}\}}_{j,k,\ell}.\end{equation}  The
          probability, $P^{\{\psi_{n}\}}_{\ell}$, that there exists an
          $h$ such that the sequence $\{\psi_{n}\}$ satisfies the Cauchy
          conditions at $\ell$ for all $j,k>h$ is given by \begin{equation}
          \label{probell}P^{\{\psi_{n}\}}_{\ell}=\limsup_{h\rightarrow\infty}
          P^{\{\psi_{n}\}}_{h,\ell}.\end{equation} Finally one includes all
          $\ell$ by \begin{equation}\label{probcauchy}P^{\{\psi_{n}\}}=
          \liminf_{\ell\rightarrow\infty}P^{\{\psi_{n}\}}_{\ell}.
          \end{equation} Putting these equations together gives\begin{equation}
          \label{convfin}P^{\{\psi_{n}\}}=\liminf_{\ell\rightarrow\infty}
          \limsup_{h\rightarrow\infty}\liminf_{j,k>h}P^{\{\psi_{n}\}}_{j,k,\ell}
          \end{equation} where $P^{\{\psi_{n}\}}_{j,k,\ell}$ is given by Eq. \ref{probjkl}.

          One can use these equations to obtain necessary and
          sufficient conditions for the sequence $\{\psi_{n}\}$ to
          be Cauchy with probability $P^{\{\psi_{n}\}}=1.$ One
          condition is that $P^{\{\psi_{n}\}}_{\ell}=1$ for each $\ell.$ This
          follows from Eq. \ref{probcauchy} which shows that
          $P^{\{\psi_{n}\}}$ is the greatest lower bound of all the
          $P^{\{\psi_{n}\}}_{\ell}.$  This condition is satisfied if, for each $\ell,$
          $P^{\{\psi_{n}\}}_{h,\ell}$ either equals $1$ for all $h$ greater
          than some $h_{0}$ or approaches $1$ asymptotically as
          $h\rightarrow\infty.$ This follows from Eq.
          \ref{probell} which gives $P^{\{\psi_{n}\}}_{\ell}$ as the least
          upper bound of all the $P^{\{\psi_{n}\}}_{h,\ell}.$ These two
          conditions can be combined to the single condition that
          for all $\ell,$ $P^{\{\psi_{n}\}}_{j,k,\ell}\rightarrow 1$ as
          $j,k\rightarrow\infty$ or
          $\liminf_{j,k\rightarrow\infty}P^{\{\psi_{n}\}}_{j,k,\ell}=1.$

          It is useful at this point to consider examples.
          Let $s$  be a $0-1$ valued function on the infinite
          integer interval $[0,-\infty]$  and let the sign qubits be
          at site $m=0$ and in state $+$.  Define each $\psi_{n}$ in
          the sequence $\{\psi_{n}\}$   by\begin{equation}\label{exam1}
          \psi_{n} =\cd_{+,0}\ad_{s(0),0}\cdots\ad_{s(-n+1),-n+1}
          \frac{1}{\sqrt{2}}(\ad_{1,-n}+\ad_{0,-n}|0\rangle.
          \end{equation}  This is a simple example of a pure real
          (no imaginary component) state sequence that does
          not correspond to any classical complex rational number sequence.
          The observation that the probability is $1$ that this
          sequence is Cauchy follows from the fact that for each
          $\ell,$ the probability $P^{\{\psi_{n}\}}_{j,k,\ell} =1$ for all $j,k>\ell.$
          It follows that $P^{\{\psi_{n}\}}_{\ell}=1$ for each
          $\ell$ and thus $P^{\{\psi_{n}\}}=1.$

          There are many simple examples of this type. For instance,
          one can include an imaginary component to $\psi_{n}$ by
          letting $t$ be a $0-1$ valued function with the same
          domain as $s,$ including a string of $\dd_{+,0}\bd_{t(0),0}
          \cdots\bd_{t(-n+1),-n+1},$ of creation operators, and
          replacing the superposition state at site $-n$ by a
          Bell state operator\begin{equation}\label{defBn}
          B_{-n}=\frac{1}{\sqrt{2}}(\ad_{1,-n}\bd_{1,-n}+
          \ad_{0,-n}\bd_{0,-n}).\end{equation}$\{\psi_{n}\}$ is
          still a Cauchy sequence even though the component states
          of the sequence are entangled.

          There are also more complex examples of Cauchy
          sequences based on rational approximations to analytical
          functions. An example of this type is based on a rational
          approximation to a Gaussian function. Let $s$ be a $0-1$
          valued function on the infinite interval domain
          $[u,-\infty]$ with the sign at $m=0$.  Define
          $|+,S(s',n)\rangle$ to be the $nth$ Gaussian approximation
          to  the state $|+,s'\rangle.$  That is,\begin{equation}\label{exam2}
          |+,S(s',n)\rangle =_{A}|(\exp-[\frac{|((+,
          s')-_{A}(+,s_{[u,-n]}))^{2}\rangle\times_{A} |+,n\rangle}
          {|+,\sigma\rangle^{2}}]_{n})_{n}\rangle.\end{equation}
          Here  $|((+,s')-_{A}(+,s_{[u,-n]}))^{2}\rangle$ is the
          state resulting from subtracting $|+,s_{[u,-n]}\rangle$
          from $|+,s'\rangle$ and multiplying the result by itself.
          $|+,n\rangle$ is a natural number state, and $|+,\sigma\rangle$
          is a positive rational state. No imaginary components are present.
          The subscripts $"n"$ on $[-]$ and $(-)$ denote division to accuracy $n$
          and evaluation of the exponential to accuracy $n$, perhaps
          as an initial part of a power series expansion. Note that
          $n$ appears both in the exponent and as accuracy subscripts.

          Define the state $\psi_{n}$ by \begin{equation}\label{exam2state}
          \psi_{n}=\sum_{s'}^{n}\frac{\langle +,S(s',)|\tilde{N}
          |+,S(s',n)\rangle}{M_{n}}|+,s'\rangle.\end{equation}  Here
          the matrix element is the $\tilde{N}$ eigenvalue
          associated with the state $|+,S(s',n)\rangle$ and
          $\tilde{N}$ is given by Eqs. \ref{def1N}-\ref{NXv}.

          The superscript $n$ on the summation means the sum is
          restricted to all $s$ with a domain $[u+n,-n]$.  The
          restriction to a finite domain is necessary because of
          the presence of states with arbitrary numbers of leading
          and trailing $0s.$ Without such a restriction it would be
          difficult, if not impossible to normalize $\psi_{n}$ with
          some normalization factor, $M_{n}$.

          The coefficients on the right side of Eq. \ref{exam2}
          correspond to a rational approximation to accuracy
          $n$ of a Gaussian distribution about $|+,s_{[u,-n]}\rangle$
          (or about the eigenvalue $N(+,s_{[u,-n]}).$ The presence of
          $n$ in the numerator of the exponent ensures that the sequence
          $\{\psi_{n}\}$ is Cauchy. This follows from the
          observation that the standard deviation, $\sigma^{2}_{n}=
          \sigma^{2}/n\rightarrow 0$ as $n\rightarrow\infty.$

          \section{Properties of and Operations on Cauchy
          State Sequences}\label{POCSS}
           Cauchy state sequences inherit many properties of
          the complex rational string states.  They also have some
          additional properties that are not possessed by the string
          states. Here the emphasis is on properties and
          operations needed to show that (equivalence classes of)
          Cauchy state sequences have the requisite properties
          of real and complex numbers. The basic relations are
          equality $=_{X}$ and an ordering $<_{X}$ for $X=R,I,$ and
          $C$. These refer to equality  and ordering defined separately
          for the real and imaginary components and for both together.
          As is well known, $<_{C}$ is defined only on those complex
          state pairs where both the real and imaginary components
          have the same ordering relation.

          The basic operations on Cauchy sequences are those of a field,
          namely, $\tilde{+}_{X},\tilde{-}_{X},\tilde{\times}_{X},$ and
          $\tilde{\div}_{X}.$ The definitions of these operators
          will follow those for the arithmetic operations in that
          their action on pairs of Cauchy sequence leaves the pairs
          and creates a third sequence of states. The actions  of
          these operators on Cauchy sequences of rational string states,
          can be represented by\begin{equation}\label{OpsCau}\begin{array}{l}
          \tilde{O}_{X}\{|\g_{n}s_{n};(\g_{1})_{n}t_{n}\rangle\}
         \{|\gp_{n}\sp_{n};(\g'_{1})_{n}t'_{n}\rangle\}\\ \hspace{0.5cm}
          =\{|\g_{n}s_{n};(\g_{1})_{n}t_{n}\rangle\}\{|\gp_{n}\sp_{n};
          (\g'_{1})_{n}t'_{n}\rangle\}\{|\g''_{n}s''_{n};
          (\g_{1}'')_{n}t''_{n}\rangle\}\end{array}\end{equation}
          where $O$ stands for $+,-,\times,$ and $\div.$ Here $\{|\g''_{n}s''_{n};
          (\g_{1}'')_{n}t''_{n}\rangle\}$ is the sequence resulting
          from carrying out the operation $O$.

          The state sequence, $\{|\g''_{n}s''_{n};(\g_{1}'')_{n}t''_{n}
          \rangle\},$ which is the result of carrying out the
          operation $\tilde{O}_{X},$ is defined by\begin{equation}\label{opres}
          \{|\g''_{n}s''_{n};(\g_{1}'')_{n}t''_{n}\rangle\}=\{|(\g_{n}s_{n};
          (\g_{1})_{n}t_{n})O_{A}(\g'_{n}s'_{n};(\g'_{1})_{n}t'_{n})\rangle.\}
          \end{equation} For each $n$ the $nth$ element of this
          sequence is the state obtained by carrying out the
          arithmetic $O_{A}$ operation on the $nth$ elements of the pair of
          input Cauchy sequences.

          This definition is satisfactory for all operations except division
          as the string states are not closed under division. One
          way around this is to use a diagonal definition:  The
          $nth$ element of$\{|\g''_{n}s''_{n};(\g_{1}'')_{n}
          t''_{n}\rangle\}$ is defined by $|(\g_{n}s_{n};
          (\g_{1})_{n}t_{n})\div_{A,n}(\g'_{n}s'_{n};(\g'_{1})_{n}
          t'_{n})\rangle.$ More details will be given later.

          Note that the definitions of both the properties and operations
          are global in that they apply to tuples of state sequences
          anywhere in $I\times I.$ This is implicitly assumed
          although it could be made explicit by including the
          location parameters and summing over them with the
          restriction that no two sequences have the same $h$ value.

          In the following the definitions of the properties and
          operations are extended to sequences of linear
          superposition states.  Also, proofs that the resulting
          state sequences satisfying Eq. \ref{opres} are Cauchy are
          provided.

          \subsection{The Properties $=_{X}$ and $<_{X}$ for $X=R,I,C$}

           A first step is to lift the properties $=_{A}$ and
           $<_{A}$ from states in $\mathcal F$ to Cauchy
           sequences of these states. Two Cauchy sequences of
           real rational states, $\{|\g_{n},s_{n}\rangle\}$ and
          $\{|\gp_{n},\sp_{n}\rangle\}$, are $R$ equal,
          \begin{equation}\label{seqR=}\{|\g_{n},s_{n}
          \rangle\}=_{R}\{|\gp_{n},\sp_{n}\rangle\}
          \end{equation} if  for all $\ell$ there is an $h$ such
           that for all $j,k>h$ \begin{equation}\label{cauN=}
          |(|\g_{j}s_{j}-_{A}\gp_{k}\sp_{k}|_{A})\rangle \leq_{A}|+,-\ell\rangle.
           \end{equation} Cauchy sequences of complex rational
          states,  $\{|\g_{n},s_{n},(\g_{1})_{n},t_{n}\rangle\}$
          and $\{|\gp_{n},\sp_{n},(\gp_{1})_{n}
          \tp_{n}\rangle\},$  are $C$ equal
          \begin{equation}\label{seqC=}\{|\g_{n},s_{n},(\g_{1})_{n},t_{n}
          \rangle\}=_{C}\{|\gp_{n},\sp_{n},(\gp_{1})_{n},\tp_{n}\rangle\}
          \end{equation} if for all $\ell$\begin{equation}
          \label{cauN=ri}\begin{array}{c}
          |(|\g_{j},s_{j}-_{A}\gp_{k},\sp_{k}|_{A})\rangle \leq_{A}|+,-\ell\rangle
          \\ |(|(\g_{1})_{j},t_{j}-_{A}(\gp_{1})_{k},\tp_{k}|_{A})\rangle
          \leq_{A}|+,-\ell\rangle\end{array}\end{equation} for all $j,k>$ some $h.$
           Note that $=_{C}$ is equivalent to both
           $=_{R}$ and $=_{I}$ holding for the real and imaginary parts.
           One can also have Cauchy sequences in which either one, but not
           both, of the real and imaginary parts are  equal.

            Extension of these definitions to
           sequences of states that are linear superpositions of
           rational string states is based on the results obtained so
           far. The Cauchy sequences, $\{\psi_{n}\}$ and $\{\psi^{\p}_{n}\},$
           are $C$ equal, \begin{equation}\label{cauchyseq=}
          \{\psi_{n}\}=_{C}\{\psi^{\p}_{n}\}\end{equation} if
          $\liminf_{j,k\rightarrow\infty}P_{j,k,\ell} =1$ where
          the probability $P_{j,k,\ell}$ is given by Eq. \ref{probjkl}
          with $\psi^{\p}_{k}$ replacing $\psi_{k}.$

           As would be expected, there are many different Cauchy
           sequences that are $C$ equal to a given sequence.  It is also the case
          that for each Cauchy sequence $\{\psi_{n}\}$ there is a Cauchy
          sequence,  $\{|\g_{n}, s_{n},\g'_{n},t_{n}\rangle\},$ of complex
          rational states that is $C$ equal to $\{\psi_{n}\}.$
          The proof or this is first given for real states and then extended
          to complex states. For real states the proof requires finding a
          rational string state sequence $\{|\g_{n},s_{n}\rangle\}$
          where the probability is one that $\{|\g_{n},s_{n}\rangle\}=_{R}
          \{\psi_{n}\}.$

          The probability condition can be expressed by first
          defining $Q_{j,\ell,\gp,\sp}$ for any $|\gp,\sp\rangle$ by
          \begin{equation}\label{defpjkl}Q_{j,\ell,\gp,\sp}=
          \sum_{\g,s}|\langle\g,s|\psi_{j}\rangle|^{2}:|(|\g,
          s-_{A}\gp,\sp|_{A})\rangle<_{A}|+,-\ell\rangle.\end{equation}
          Comparison with Eq. \ref{probjkl}, restricted to real rational states,
          shows that\begin{equation}\label{PQjkl}P_{j,k,\ell}=\sum_{\gp,\sp}
          |\langle\gp,\sp|\psi_{k}\rangle|^{2}Q_{j,\ell,\gp,\sp}.\end{equation}

          For each $j$ define  $|\g_{j},s_{j}\rangle$ to be the
          string state $|\gp, \sp\rangle$ that maximizes
          $Q_{j,\ell,\gp,\sp}.$  Let $Q_{j,\ell,\g_{j},s_{j}}$
          be the maximum value. One sees immediately that
         \begin{equation}\label{P<Q} P_{j,k,\ell}\leq \sum_{\gp,\sp}
          |\langle\gp,\sp|\psi_{k}\rangle|^{2}Q_{j,\ell,\g_{j},s_{j}}=
          Q_{j,\ell,\g_{j},s_{j}}.\end{equation} Since
          $\{\psi_{n}\}$ is Cauchy, $P_{j,k,\ell}\rightarrow 1$ as
          $j,k,\rightarrow\infty$ which gives
          $Q_{j,\ell,\g_{j}s_{j}}\rightarrow 1$ as $j\rightarrow\infty$
          for each $\ell.$ This completes the  proof that
          $\{\psi_{n}\}=_{R}\{|\g_{n}s_{n}\rangle\}.$

           The proof for complex states follows that
           already given.  One must show that for any Cauchy
           sequence $\{\psi_{n}\},$ there is a Cauchy
          sequence $\{|\g_{n},s_{n},\gp_{n},t_{n}\rangle\}$ where
          $\{|\g_{n},s_{n},\gp_{n},t_{n}\rangle\}=_{C}\{\psi_{n}\}.$
          Following Eq. \ref{defpjkl} one defines for any
          $|\g,s,\gp,t\rangle$ \begin{equation}
          \label{equivricauch} \begin{array}{l}Q_{j,\ell,\g,s,\gp,t}=
          \sum_{\g,s,\gp,t}|\langle\g, s,\gp, t|\psi_{j}\rangle|^{2}:\\
          \hspace{1cm}|(|\g, s-_{A}\g_{1},\sp|_{A})\rangle <_{A}|+,-\ell\rangle
          \\\hspace{1cm}\mbox{ and } |(|\g, t-_{A}\gp_{1},\tp|_{A})\rangle
          <_{A}|+,-\ell\rangle.\end{array}\end{equation}
          For each $j$ define $|\g_{j},s_{j},\gp_{j},t_{j}\rangle$ to
          be the complex rational string state that maximizes
          $Q_{j,\ell,\g,s,\gp,t}.$ Then Eq. \ref{probjkl} gives
          \begin{equation}\label{P<Qri}\begin{array}{l}
          P_{j,k,\ell}\leq \sum_{\g,s,\gp,t}
          |\langle\g,s,\gp,t|\psi_{k}\rangle|^{2}Q_{j,\ell,
          \g_{j},s_{j},\gp_{j},t_{j}}\\ \hspace{1.5cm}=
          Q_{j,\ell,\g_{j},s_{j},\gp_{j},t_{j}}.\end{array}\end{equation}
          Since $\{\psi_{n}\}$ is Cauchy, $P_{j,k,\ell}\rightarrow 1$ as
          $j,k\rightarrow\infty.$ This implies that
          $ Q_{j,\ell,\g_{j},s_{j},\gp_{j},t_{j}}\rightarrow 1 $ as
          $j,k\rightarrow\infty,$ which completes the proof.

          Definitions of $<_{R},<_{I},$ and $<_{C}$ on Cauchy sequences of
          rational states  are based on the definition of $<_{A}.$
          The Cauchy sequence $\{|\g_{n},s_{n}\rangle\}$ of
          real rational states is $R$ less than $\{|\gp_{n},\sp_{n}\rangle\},$
          \begin{equation}\label{seqR<}\{|\g_{n},s_{n}
          \rangle\}<_{R}\{|\gp_{n},\sp_{n}\rangle\}\end{equation}
          if for some $\ell$ and $h$ \begin{equation}\label{def<R}
           |(\g_{j},s_{j})+_{A}(+,-\ell)\rangle
          <_{A}|\gp_{k},\sp_{k}\rangle\end{equation} for all $j,k>h.$ This
          is based on the observation that two Cauchy sequences
          are not $R$ equal if they are separated asymptotically
          by a finite gap, denoted by  $(+,-\ell)$ in the state on
          the left.

          A similar definition of $<_{C}$ applies to  Cauchy
          sequences of complex rational states. \begin{equation}\label{<Cdef}
          \{|\g_{n},s_{n}\gp_{n}t_{n}\rangle\}<_{C}
          \{|(\g_{1})_{n},\sp_{n},(\gp_{1})_{n},\tp_{n}\rangle\}
          \end{equation} if both the real and imaginary
          component sequences are separated asymptotically by
          gaps. Of course $<_{C}$ is only partially defined on
          these sequences as the real and imaginary parts of a
          sequence can have different order relations.

         The ordering relations $<_{R},$ $<_{I}$, and $<_{C}$
         can be extended to Cauchy sequences of superposition states.
         One has \begin{equation}\label{ps<Npsp}
          \{\psi_{n}\}<_{R}\{\psi^{\p}_{n}\}\end{equation} with
          probability $1$ if for some $|+,-\ell\rangle$,
          $\lim_{j,k\rightarrow\infty}Q_{R,j,k,\ell}=1$ where
          \begin{equation}\label{seq<N}\begin{array}{l}
          Q_{R,j,k,\ell}=\sum_{\g,s}\sum_{\g_{1},\sp}
          |\langle\g,s|\psi_{j}\rangle|^{2}\\ \times
          |\langle\g_{1},\sp|\psi^{\p}_{k}\rangle|^{2}:
          |(\g, s)+_{A}(+,-\ell)\rangle <_{A}|\g_{1},\sp\rangle.\end{array}
          \end{equation}  That is, the probability is $1$ that
          the real parts of $\{\psi_{n}\}$ and
          $\{\psi^{\p}_{n}\}_{n}$ are separated asymptotically by
          a gap.  Similar relations hold for $<_{I}$ and $<_{C}$.

          Sequence pairs $\{\psi_{n}\}$ and $\{\psi'_{n}\}$
          that are Cauchy satisfy the following relations for
          $<_{X}$ and $=_{X}$ for $X=R$ or $X=I,$
          \begin{equation}\label{<=>psi}\begin{array}{l}\{\psi_{n}\}<_{X}
          \{\psi^{\p}_{n}\} \mbox { true with probability $1$ or} \\
          \{\psi_{n}\}=_{X}\{\psi^{\p}_{n}\}\mbox{ true with
          probability $1$ or} \\ \{\psi^{\p}_{n}\}<_{X}
          \{\psi_{n}\}\mbox{ true with probability $1$}\end{array}
          \end{equation}  One way to show this is to prove it for
          Cauchy sequences,  $\{|\g_{n},s_{n},\gp_{n},t_{n}\rangle\}$,
          of complex rational string states and use the fact that any Cauchy
          sequence $\{\psi_{n}\}$ is $C$ equal to some such sequence.

         Eq. \ref{<=>psi} does not hold in general for $X=C.$ An
         example  would be a pair of Cauchy sequences
         in which the real and imaginary parts satisfy different
         alternatives in the equation, such as
         $\{\psi_{n}\}<_{R}\{\psi^{\p}_{n}\}$ and
         $\{\psi_{n}\}>_{I}\{\psi^{\p}_{n}\}.$

         \subsection{Addition and Subtraction}

          As shown by Eqs. \ref{OpsCau} and \ref{opres}, addition of
          two Cauchy sequences of rational
          states, $\{|\g_{n},s_{n},(\g_{1})_{n},t_{n}\rangle\}$ and
          $\{|\gp_{n},\sp_{n},(\gp_{1})_{n},\tp_{n}\rangle\}$, gives the
          state sequence $\{|\g_{n}, s_{n}+_{A}\gp_{n},\sp_{n},
          (\g_{1})_{n},t_{n}+_{A}(\gp_{1})_{n},\tp_{n}\rangle\}.$
          Proof that this sequence is Cauchy requires showing
          that for all $\ell$ there is an $h$ such that
          \begin{equation}\label{+cauchy}\begin{array}{l}|(|\g_{j},
          s_{j}+_{A}\gp_{j},\sp_{j})-_{A}(\g_{k},
          s_{k}+_{A}\gp_{k},\sp_{k}|_{A})\rangle<_{A}|+,-\ell\rangle
          \mbox{ and } \\ |(|(\g_{1})_{j},t_{j}+_{A}(\gp_{1})_{j},\tp_{j})-_{A}
          ((\g_{1})_{k},t_{k}+_{A}(\gp_{1})_{k},\tp_{k}|_{A})
          \rangle<_{A}|+,-\ell\rangle\end{array}
          \end{equation} for all $j,k>h.$ Rearranging the terms in
          the left hand parts of the inequalities and using
          \begin{equation}\label{ineqabs}\begin{array}{l}
          |(|\g_{j},s_{j}-\g_{k},s_{k}+\gp_{j},\sp_{j}-\gp_{k},\sp_{k}|)\rangle
          \\ \hspace{.5cm}<_{A}|(|\g_{j},s_{j}-\g_{k},s_{k})|\rangle +_{A}|
          (|\gp_{j},\sp_{j}-_{A}\gp_{k},\sp_{k}|_{A})\rangle; \\ |(|(\g_{1})_{j},
          t_{j}-_{A}(\g_{1})_{k},t_{k}+_{A}(\gp_{1})_{j},\tp_{j}-_{A}(\gp_{1})_{k},
          \tp_{k}|_{A})\rangle \\ \hspace{.5cm}<_{A}
          |(|(\g_{1})_{j},t_{j}-_{A}(\g_{1})_{k},t_{k}|_{A})\rangle +_{A}
          |(|(\gp_{1})_{j},\tp_{j}-_{A}(\gp_{1})_{k},\tp_{k}|_{A})\rangle,\end{array}
          \end{equation} gives Eq. \ref{+cauchy} with $\ell$
          replaced by $\ell-1.$ This result uses the Cauchy
          property of the two sequences
          $\{|\g_{n},s_{n},(\g_{1})_{n},t_{n}\rangle\}$ and
          $\{|\gp_{n},\sp_{n},(\gp_{1})_{n},\tp_{n}\rangle\}.$
          Eq. \ref{++plus} was used to equate $|+,-\ell\rangle
          +_{A}|+,-\ell\rangle$ to $|+,-(\ell-1)\rangle.$

          Addition\footnote{From now
          on the subscript $A$ will not be used when it is clear
          that the relations are arithmetic.} of two Cauchy sequences $\{\psi_{n}\},
          \{\psi^{\p}_{n}\}$ gives the sequence of density operator
          states $\{\rho_{\psi_{n}+\psi^{\p}_{n}}\}$ where by Eq.
          \ref{rhoplus}
          \begin{equation}\label{rhoplusseq}\begin{array}{l}
        \rho_{\psi_{n}+\psi^{\p}_{n}}=\sum_{\g,s,\g_{1},t}\sum_{\gp, \sp,\gp_{1},\tp}
        |\langle\g,s,\g_{1},t|\psi_{n}\rangle|^{2}\\ \hspace{1cm}
        |\langle\gp,\sp,\gp_{1},\tp|\psi^{\p}_{n}\rangle|^{2}
        \rho_{(\g,s,\g_{1},t)+(\gp,\sp,\gp_{1},\tp)} \end{array}\end{equation} and
        $\rho_{(\g,s,\g_{1},t)+(\gp,\sp,\gp_{1},\tp)}=
        |(\g,s,\g_{1},t)+(\gp,\sp,\gp_{1},\tp)\rangle
        \langle(\g,s,\g_{1},t)+(\gp,\sp,\gp_{1},\tp)|.$

        To show that $\{\rho_{\psi_{n}+\psi^{\p}_{n}}\}$ is
        Cauchy, let $Q^{\p}_{j,k,\ell}$ be the probability  that
        $|Re\rho_{j}-Re\rho_{k}|<_{N}\rho_{+,-\ell}$ and
        $|Im\rho_{j}-Im\rho_{k}|<_{N}\rho_{+,-\ell}$ where $\rho_{j} =
        \rho_{\psi_{j}+\psi^{\p}_{j}},\; \rho_{k} =
        \rho_{\psi_{k}+\psi^{\p}_{k}},$ and
        $\rho_{+,-\ell}=|+,-\ell\rangle\langle+,-\ell|.$  This is
        given by \begin{equation}\label{Qpjkell}
        \begin{array}{l}Q^{\p}_{j,k,\ell}=\sum_{\g_{j},s_{j},(\g_{1})_{j},t_{j}}
        \sum_{\gp_{j},\sp_{j},(\gp_{1})_{j},\tp_{j}}
        \sum_{\g_{k},s_{k},(\g_{1})_{k},t_{k}}\sum_{\gp_{k},\sp_{k},(\gp_{1})_{k},\tp_{k}}
        \\ \hspace{.5cm}\times|\langle\g_{j},s_{j},(\g_{1})_{j},t_{j}|\psi_{j}\rangle|^{2}
        |\langle\gp_{j},\sp_{j},(\gp_{1})_{j},\tp_{j}|\psi^{\p}_{j}\rangle|^{2}\\
        \hspace{.5cm}\times|\langle\g_{k},s_{k},(\g_{1})_{k},t_{k}
        |\psi_{k}\rangle|^{2}|\langle\gp_{k},\sp_{k},(\gp_{1})_{k},\tp_{k}|
        \psi^{\p}_{k}\rangle|^{2}: \\ \hspace{1cm}\rho_{|\g_{j},s_{j}+\gp_{j},
        \sp_{j}-\g_{k},s_{k}-\gp_{k},\sp_{k}|} <_{A}\rho_{+,-\ell} \\ \hspace{1cm}
        \mbox{and }\rho_{(\g_{1})_{j},t_{j}+(\gp_{1})_{j},\tp_{j}-
        (\g_{1})_{k},t_{k}-(\gp_{1})_{k},\tp_{k}|}
        <_{A}\rho_{+,-\ell}.\end{array} \end{equation} The $<_{A}$ condition
        stated for the density operators is equivalent to that given by
        Eq. \ref{+cauchy} for string states.

        Let $|\g_{j},s_{j},(\g_{1})_{j},t_{j}\rangle,|\g_{k},s_{k},(\g_{1})_{k},
        t_{k}\rangle$ and $|\gp_{j},\sp_{j},(\gp_{1})_{j},\tp_{j}\rangle,
        |\gp_{k},\sp_{k},(\gp_{1})_{k},\tp_{k}\rangle$ be four string
        states that satisfy the Cauchy conditions given in Eq.
        \ref{probjkl} for $P^{\{\psi_{n}\}}_{j,k,\ell}$ and
        $P^{\{\psi^{\p}_{n}\}}_{j,k,\ell}.$ From Eq. \ref{ineqabs} one
        sees that these states also satisfy the Cauchy conditions in
        Eq. \ref{Qpjkell} for $\ell-1.$  This gives the result
        that $Q^{\p}_{j,k,\ell}$ is related to
        $P^{\{\psi_{n}\}}_{j,k,\ell}$ and $P^{\{\psi^{\p}_{n}\}}_{j,k,\ell}$ by
         \begin{equation}\label{QQp}Q^{\p}_{j,k,\ell-1}\geq
        P^{\psi}_{j,k,\ell}P^{\psi^{\p}}_{j,k,\ell}.\end{equation}
        Since  the sequences $\{\psi_{n}\}$ and $\{\psi^{\p}_{n}\}$ are Cauchy,
        $P^{\{\psi_{n}\}}_{j,k,\ell}\rightarrow 1$ and $P^{\{\psi^{\p}_{n}\}}_{j,k,\ell}
        \rightarrow 1$ as $j,k\rightarrow\infty$ for any $\ell.$  It follows that
        $Q^{\p}_{j,k,\ell-1}\rightarrow 1$ as
        $j,k\rightarrow\infty.$  It follows immediately from this that
        $\{\rho_{\psi_{n}+_{A}\psi^{\p}_{n}}\}$  is a Cauchy sequence.

        Based on the results obtained so far, other properties of addition
        of Cauchy state sequences can be proved.  These include
        commutativity, associativity, and any sequence $\{\psi_{n}\}$
        which converges to $0$ (or $\{\psi_{n}\}=_{C}\{|+,0\rangle_{n}\},$
        the constant $0$ state sequence), is an additive
        identity, etc. The definition of subtraction, as the
        inverse of addition, is  straightforward as
        \begin{equation}\label{seqsubtr}\begin{array}{l}\{|\g_{n},
        s_{n},(\g_{1})_{n},t_{n}-\gp_{n},\sp_{n},(\gp_{1})_{n},\tp_{n}\rangle\}
        =_{C} \\ \hspace{0.5cm}\{|\g_{n},s_{n},(\g_{1})_{n},t_{n}
        +\g^{\p\p}_{n},\sp_{n},(\g^{\p\p}_{1})_{n},\tp_{n}\rangle\}.
        \end{array}\end{equation} Here $\g^{\p\p}_{n}\neq \gp_{n}$
        and $(\g^{\p\p}_{1})_{n}\neq (\gp_{1})_{n}.$

        \subsection{Multiplication and Division}

        For multiplication  it is useful to first consider sequences
        of real rational states and then extend the results to  complex
        rational state sequences. The goal is to show that the
        product state sequence, $\{|\g_{n}s_{n}\times\gp_{n}
        \sp_{n}\rangle\},$ of two Cauchy  sequences,
        $\{|\g_{n}s_{n}\rangle\}$ and
        $\{|\gp_{n}\sp_{n}\rangle\},$ is a Cauchy sequence.
        For all $j,k>$ some $h$, \begin{equation}\label{mult<N}
        \begin{array}{l} |(|\g_{j}s_{j}\times\gp_{j}\sp_{j}-\g_{k}s_{k}\times
        \gp_{k}\sp_{k}|)\rangle \\\leq_{A}|(|\g_{j}s_{j}-_{A}\g_{k}s_{k}
        |_{A}\times_{A}|\gp_{j}\sp_{j}|_{A}
        +_{A}|\g_{k}s_{k}|_{A}\times_{A}|\gp_{j}\sp_{j}-_{A}\gp_{k}\sp_{k}
        |_{A})\rangle \\\hspace{1cm} <_{A}|(|\g_{j}s_{j}-_{A} \g_{k}s_{k}|_{A}+
        |\gp_{j}\sp_{j}-_{A} \gp_{k}\sp_{k}|_{A})\times_{A}(+,\ell_{u})
        \rangle \\ \hspace{1.5cm}<_{A}|+,-(\ell-\ell_{u})\rangle.\end{array}
        \end{equation} Here $|+,\ell_{u}\rangle$ is an upper bound to
        $|(|\gp_{j}\sp_{j}|_{A})\rangle$ and to $|(|\g_{k}s_{k}|_{A})\rangle$
        for all $j,k.$ Such a bound exists because  $\{|\g_{n}s_{n}\rangle\}$ and
        $\{|\gp_{n}\sp_{n}\rangle\}$ are Cauchy sequences. Since
        $\ell_{u}$ is fixed and is independent of $\ell,$ Eq. \ref{mult<N}
        shows that $|(|\g_{j}s_{j}\times\gp_{j}\sp_{j}-\g_{k}s_{k}\times
        \gp_{k}\sp_{k}|)\rangle\rightarrow |0\rangle$ as $j,k\rightarrow\infty.$
        This shows that the product sequence $\{|\g_{n}s_{n}\times
        \gp_{n}\sp_{n}\rangle\}$ is Cauchy.

        This result can be extended directly to products of
        Cauchy sequences of complex rational states.   Let
        $\{|\g_{n}s_{n},(\g_{1})_{n}t_{n}\rangle\}$ and
        $\{|\gp_{n}\sp_{n},(\gp_{1})_{n}\tp_{n}\rangle\}$ be two
        Cauchy sequences. The product sequence,
        $\{\g_{n}s_{n},(\g_{1})_{n}t_{n}\times\gp_{n}\sp_{n}
        (\gp_{1})_{n}\tp_{n}\rangle\},$ with the real and
        imaginary parts separated, is given by
        $\{|[(\g_{n}s_{s}\times\gp_{n}\sp_{n})+((\g_{1})_{n}t_{n}\times
        (\gp_{1})_{n}\tp_{n})],[(\g_{n}s_{s}\times(\gp_{1})_{n}\tp_{n})+
        (\gp_{n}\sp_{n}\times(\g_{1})_{n}t_{n})]\rangle\}.$ To save on
        notation let this sequence be represented by
        $\{|\eta_{n}v_{n},\delta_{n}w_{n}\rangle\}$ where
        $\eta_{n}v_{n}$ is the real part and $\delta_{n}w_{n}$ is
        the imaginary part. To prove that the product sequence is
        Cauchy. it is sufficient to show that
        \begin{equation}\label{<Nreim}
        \begin{array}{l} |(|\eta_{j}v_{j}-\eta_{k}v_{k}|)\rangle
        <_{A}|+,-\ell^{\p}\rangle \\ |(|\delta_{j}w_{j}-\delta_{k}w_{k}|)
        \rangle<_{A}|+,-\ell^{\p}\rangle,\end{array}\end{equation}
        then \begin{equation}\label{prodric}|(|\eta_{j}v_{j}+\delta_{j}w_{j}-
        \eta_{k}v_{k}-\delta_{k}w_{k}|)\rangle<_{A}|+,-(\ell^{\p}-1)\rangle.
        \end{equation}

        To prove Eq. \ref{<Nreim} one has \begin{equation}\label{ab<N}
        \begin{array}{l}|(|\eta_{j}v_{j}-\eta_{k}v_{k}|)\rangle\leq_{A}
        |(|\g_{j}s_{j}\times\gp_{j}\sp_{j}-\g_{k}s_{k}\times
        \gp_{k}\sp_{k}|)\rangle \\ \hspace{1cm} +_{A}  |(|(\g_{1})_{j}t_{j}\times
        (\gp_{1})_{j}\tp_{j}-(\g_{1})_{k}t_{k}\times
        (\gp_{1})_{k}\tp_{k}|)\rangle.\end{array}
        \end{equation} Applying the argument used to verify Eq. \ref{mult<N} to
        each state gives\begin{equation}\label{gammavell}
        |(|\eta_{j}v_{j}-\eta_{k}v_{k}|)\rangle
        <_{A}|+,-(\ell-1-\ell_{u})\rangle.\end{equation}  Here
        $|+,\ell_{u}\rangle$ is an upper bound to
        $|(|\g_{j}s_{j}|)\rangle,|(|\gp_{j}\sp_{j}|)\rangle,|(|(\g_{1})_{j}t_{j}
        |)\rangle,|(|(\gp_{1})_{j}\tp_{j}|)\rangle$ for all $j$. Applying
        the same argument to $|(|\delta_{j}w_{j}-\delta_{k}w_{k}|)
        \rangle$ and setting $\ell^{\p}=\ell-1-\ell_{u}$ finishes
        the proof.

        For Cauchy sequences, $\{\psi_{n}\}$ and
        $\{\psi^{\p}_{n}\},$ of states that are linear
        superpositions of complex string states,
        the product states $\rho_{\psi_{n}\times\psi^{\p}_{n}}$
        in the sequence of density operators are given by
        Eq. \ref{rhoplusseq} with $\rho_{(\g,s,\g_{1},t)\times(\g,s,\g_{1},tp)}$
        replacing $\rho_{(\g,s,\g_{1},t)+(\g,s,\g_{1},tp)}$ on the right hand side.

        To prove that $\rho_{\psi_{n}\times\psi^{\p}_{n}}$ is Cauchy, it is
        convenient to first suppress the
        imaginary component and  consider just the real string states.
        In this case the probability, $Q^{\times}_{j,k,\ell}$ that
        $|\rho_{j}-\rho_{k}|_{N}<_{A}\rho_{+,-\ell}$ is given by
        \begin{equation}\label{Qtimesjkl}
        \begin{array}{l}Q^{\times}_{j,k,\ell}=\sum_{\g_{j}s_{j}}
        \sum_{\gp_{j}\sp_{j}}
        \sum_{\g_{k}s_{k}}\sum_{\gp_{k}\sp_{k}} \\
        \times |\langle\g_{j}s_{j}|
        \psi_{j}\rangle|^{2}|\langle\gp_{j}\sp_{j}|\psi^{\p}_{j}\rangle|^{2}
        |\langle\g_{k}s_{k}|\psi_{k}\rangle|^{2}|\langle\gp_{k}\sp_{k}|
        \psi^{\p}_{k}\rangle|^{2}: \\ \hspace{1cm}\rho_{|\g_{j}s_{j}\times\gp_{j}
        \sp_{j}-\g_{k}s_{k}\times\gp_{k}\sp_{k}|}
        <_{A}\rho_{+,-\ell}.\end{array} \end{equation} The condition on
        the density operators is  equivalent to the condition
        $|(|\g_{j}s_{j}\times\gp_{j}\sp_{j}-\g_{k}s_{k}\times
        \gp_{k}\sp_{k}|)\rangle<_{A}|+,-\ell\rangle$ for the pure
        states.

        One would like to use the righthand inequality of Eq.
        \ref{mult<N} for the proof.  However there is a
        problem in that the middle inequality does not hold because
        the states $|(|\gp_{j}\sp_{j}|)\rangle$ and
        $|(|\g_{k}s_{k}|)\rangle$ have no arithmetic upper bound. However, because
        $\{\psi_{n}\}$ and $\{\psi^{\p}_{n}\}$ are Cauchy, there
        exists an $\ell_{u}$ such that the probabilities \begin{equation}
        \label{deflu}\begin{array}{l}P^{\psi_{j}}_{\ell_{u}}=\sum_{\gp_{j},s'_{j}}
        |\langle\gp_{j},s'_{j}|\psi_{j}|^{2}:|(|\gp_{j}\sp_{j}|)\rangle<_{A}|+,
        \ell_{u}\rangle \\ P^{\psi_{k}}_{\ell_{u}}=\sum_{\g_{k},s_{k}}
        |\langle\g_{k},s_{k}|\psi_{k}|^{2}: |(|\g_{k}s_{k}|)\rangle<_{A}|+,\ell_{u}
        \rangle\end{array}\end{equation} converge to $1$ as
        $j,k\rightarrow\infty.$

        Let $P^{\{\psi_{n}\}}_{j,k,\ell^{\p},\ell_{u}}$ and
        $P^{\{\psi'_{n}\}}_{j,k,\ell^{\p},\ell_{u}}$ be
        defined by \begin{equation}\label{problulp}\begin{array}{l}
        P^{\{\psi_{n}\}}_{j,k,\ell^{\p},\ell_{u}}=\sum_{\g_{j}s_{j}}
        \sum_{\g_{k}s_{k}}|\langle\g_{j}s_{j}|\psi_{j}\rangle|^{2}
        |\langle\g_{k}s_{k}|\psi_{k}\rangle|^{2}: \\ \hspace{1cm}
        |(|\g_{k}s_{k}|)\rangle<_{A}|+,\ell_{u}\rangle \\ \hspace{1cm} \mbox{ and
        }|(|\g_{j}s_{j}-\g_{k}s_{k}|)\rangle<_{A}|+,-\ell^{\p}\rangle;
        \\ P^{\{\psi'_{n}\}}_{j,k,\ell^{\p},\ell_{u}}=\sum_{\gp_{j}\sp_{j}}
        \sum_{\gp_{k}\sp_{k}}|\langle\gp_{j}\sp_{j}|\psi^{\p}_{j}
        \rangle|^{2}|\langle\gp_{k}\sp_{k}|\psi^{\p}_{k}\rangle|^{2}:
        \\ \hspace{1cm}|(|\gp_{k}\sp_{k}|)\rangle<_{A}|+,\ell_{u}\rangle
        \\ \hspace{1cm}\mbox{ and }|(|\gp_{j}\sp_{j}-\gp_{k}\sp_{k}|)\rangle<_{A}
        |+,-\ell^{\p}\rangle.\end{array}\end{equation} The Cauchy
        conditions for $\{\psi_{n}\}$ and $\{\psi^{\p}_{n}\}$ give
        the result that for some
        $\ell_{u}$, \begin{equation}\label{limjk}\begin{array}{l}
        \lim_{j,k\rightarrow\infty}P^{\{\psi_{n}\}}_{j,k,\ell^{\p},\ell_{u}}=1
        \\ \lim_{j,k\rightarrow\infty}P^{\{\psi_{n}\}}_{j,k,\ell^{\p},
        \ell_{u}}=1 \end{array}\end{equation} for each
        $\ell^{\p}.$

        Comparison of Eq. \ref{problulp} with Eq. \ref{Qtimesjkl}
        and use of Eq. \ref{mult<N} gives the result that
        \begin{equation}\label{QgeqPP}
        Q^{\times}_{j,k,\ell^{\p}-1-\ell_{u}}\geq
        P^{\{\psi_{n}\}}_{j,k,\ell^{\p},\ell_{u}}
        P^{\{\psi'_{n}\}}_{j,k,\ell^{\p},\ell_{u}}.\end{equation}
        One sees from Eq. \ref{limjk}
        that $Q^{\times}_{j,k,\ell^{\p}-1-\ell_{u}}\rightarrow 1$
        as $j,k\rightarrow\infty.$ Since $\ell_{u}$ is fixed and
        $\ell^{\p}$ is any positive integer, it follows that
        $\rho_{\psi\times\psi^{\p}}$ is Cauchy.

        Extension of this result to include multiplication of sequences
        of superposition states over complex string states is more
        involved.  It will not be given as nothing new is added.
        Sums over $\g,s,$ are expanded to sums over $\g, s,\g_{1}, t$  and
        probabilities of the form $|\langle\g,s|\psi_{n}\rangle|^{2}$
        become $|\langle\g,s,\g_{1},t|\psi_{n}\rangle|^{2}.$

        There are several well known properties that the
        definition of multiplication given here must satisfy.
        These include commutativity, distributivity over
        additivity, and the property that any sequence that is $A$ equal to
        the constant identity sequence, $\{|+,1\rangle\}_{c}=_{R}\{\cd_{+,m}
        \ad_{1,m}|0\rangle\}_{c},$ is a multiplicative identity. The subscript
        $c$ means that every element of the sequence is the same. Also if
        $\{\psi^{\p}_{n}\}=_{R}\{|+,0\rangle\}_{c}=_{R}\{\cd_{+,m}\ad_{0,m}
        |0\rangle\}_{c},$ the constant zero sequence, then for
        any Cauchy $\{\psi_{n}\},$ $\{\rho_{\psi_{n}\times\psi^{\p}_{n}}\}
        =_{C}\{\rho_{|+,0\rangle}\}_{c}.$

        Proofs of these properties for the Cauchy sequences follow the proofs of the
        Cauchy condition for the multiplicative and additive sequences. For each
        property there are conditions with associated probabilities
        of validity.  One must show that the relevant probabilities
        approach $1$ as the indices of the states in the sequences
        increase without bound. Alternatively one can prove the
        properties for Cauchy sequences $\{|\g_{n}s_{n},
        (\g_{1})_{n}t_{n}\rangle\}$  of complex string states and
        use the fact that any Cauchy sequence $\{\psi_{n}\}$
        of superposition states is $=_{C}$ to some Cauchy sequence
        $\{|\g_{n}s_{n}, (\g_{1})_{n}t_{n}\rangle\}$
        to extend the properties to the $\{\psi_{n}\}.$

        One property that should be discussed in more detail is
        the existence of a multiplicative inverse. Unlike the case for
        string states and their linear
        superpositions,  Cauchy sequences of states have
        multiplicative inverses. To see this let
        $\{|\g_{n}s_{n}\rangle\}$ be a Cauchy sequence of real
        string states where $\{|\g_{n}s_{n}\rangle\}
        \neq_{R}\{|+,0\rangle\}_{c}.$

        A sequence $\{|\gp_{n}\sp_{n}\rangle\}$ inverse to $\{|\g_{n}s_{n}
        \rangle\}$can be constructed by a diagonal process:  For each $\ell$ let
        $|\gp_{\ell}\sp_{\ell}\rangle$  be a state
        that satisfies \begin{equation}\label{divseq}
        \begin{array}{c}|(+,1)-(+,-\ell)
        \rangle\leq_{A} |\g_{\ell}s_{\ell}\times\gp_{\ell}\sp_{\ell}
        \rangle\leq_{A}|(+,1)\rangle \\ \mbox{if $|\g_{\ell}s_{\ell}
        \rangle\neq_{A}|+,0\rangle;$} \\ |\gp_{\ell}\sp_{\ell}\rangle
        =_{A}|+,1\rangle  \mbox{ if } |\g_{\ell}s_{\ell}
        \rangle =_{A}|+,0\rangle\end{array} \end{equation} This definition
        is based on the previous description, Eq. \ref{definv}, of the $\ell$
        inverse for string states.

        As noted before, for each $\ell$ and $|\g_{\ell} s_{\ell}
        \rangle\neq_{A}|+,0\rangle$,
        there are many  states  $|\gp_{\ell}\sp_{\ell}\rangle$
        satisfying Eq. \ref{divseq}. Any one of them will suffice
        here.  However a unique choice can be made by requiring
        that, of all states $|\g'' s^{\p\p}\rangle$ satisfying Eq. \ref{divseq},
        $|\gp_{\ell} \sp_{\ell}\rangle$ is the state where the
        smallest $j$ value for which $s^{\p}_{\ell}(j)=1$ is larger than that
        for any other $s^{\p\p}.$  The example following Eq. \ref{definv}
        shows how this works.

        It is clear from the definition that the
        product sequence $\{|\g_{n}s_{n}\times\gp_{n}\sp_{n}\rangle\}$
        is Cauchy and is $R$ equal to the constant unit sequence
        $\{|+,1\rangle\}_{c}.$ The Cauchy property of
        $\{|\gp_{n}\sp_{n}\rangle\}$ follows from that for
        $\{|\g_{n}s_{n}\rangle\}.$

        The construction outlined above cannot be applied directly
        to find the inverse of a Cauchy sequence
        $\{\psi_{n}\}$ of linear superposition states as
        linear superposition states do not have $\ell$ inverses.
         In this case, one is interested in finding
        for any Cauchy $\{\psi_{n}\}\neq_{C} \{|+,0\rangle\}_{c}$
         a Cauchy state sequence $\{\psi^{\p}_{n}\}$ that satisfies
       \begin{equation}\label{invpsin}\{\psi_{n}\times\psi^{\p}_{n}\}
       =_{R}\{|+,1\rangle\}_{c}.\end{equation} The meaning of this
       equation can be expressed using Eqs. \ref{probjkl}
       et.seq.  For each $j,\ell$ define the probability
       $P_{(\psi\times\psi^{\p})_{j}, \ell}$ by
       \begin{equation}\label{probpsiinv}\begin{array}{l}
       P_{(\psi\times\psi^{\p})_{j},\ell}=\sum_{\g,s}\sum_{\gp,\sp}
       |\langle\g,s|\psi_{j}\rangle|^{2}|\langle\gp,\sp|\psi^{\p}_{j}
       \rangle|^{2}: \\ \hspace{0.5cm}|(|(\g s\times\gp\sp)-(+,1)|)
       \rangle <_{A} |+,-\ell\rangle.\end{array}\end{equation}

       Eq. \ref{invpsin} is satisfied with probability one if
       \begin{equation}\label{cauchprob}\liminf_{\ell\rightarrow\infty}
       \limsup_{h\rightarrow\infty}\liminf_{j>h}P_{(\psi\times\psi^{\p})_{j},
       \ell} =1.\end{equation}  A necessary and sufficient
       condition that Eq. \ref{cauchprob} is satisfied is that
       $P_{(\psi\times\psi^{\p})_{j},\ell}\rightarrow 1$ as
       $j\rightarrow\infty$ for each $\ell.$

        The above gives the conditions to be satisfied by  a Cauchy
        sequence $\{\psi^{\p}_{n}\}$ that is inverse to  $\{\psi_{n}\}$ but it
        gives no clue as to how to construct such an inverse.
        One way to proceed is to use the substitution property of $=_{R}$
        for the property of being an inverse. (Extension to other
        properties and operations is discussed in the next section.)  Let
        $\{\psi_{n}\}$ be a Cauchy sequence of superpositions of real
        string states and $\{|\g_{n}s_{n}\rangle\}$ a Cauchy sequence
        where $\{\psi_{n}\}=_{R}\{|\g_{n}s_{n}\rangle\}.$ If
        $\{|\gp_{n}\sp_{n}\rangle\}$ is a Cauchy sequence that is the inverse of
        $\{|\g_{n}s_{n}\rangle\},$ then $\{|\gp_{n}\sp_{n}\rangle\},$
        and any Cauchy sequence $\{\psi^{\p}_{n}\}$ where
        $\{\psi^{\p}_{n}\}=_{R}\{|\gp_{n}\sp_{n}\rangle\},$ is the
        inverse of $\{\psi_{n}\}.$

        Extension of the diagonal method to construct a Cauchy
        sequence that is the inverse of the complex Cauchy sequence
        $\{|\g_{n}s_{n},(\g_{1})_{n}t_{n}\rangle\}$ is more
        complex, but nothing new is required. From Eq. \ref{definv} one has,
        for each $\ell$, \begin{equation}\label{definvcompl}
        |(+,1)-(+,-\ell)\rangle\leq_{A}
        |(+1/\g,s,\g_{1},t)_{\ell}\times_{A}(\g,s,\g_{1},t)\rangle\leq_{A}|+,1\rangle
        \end{equation} where \begin{equation}\label{defcompinv}
        |(+,1/\g,s,\g_{1},t)_{\ell}\rangle=_{A}|(+s^{\p\p})_{\ell}\times_{A}
        (\g s,\gp_{1} t)\rangle.\end{equation} Here $|(+s^{\p\p})_{\ell}\rangle$
        is given by Eq. \ref{ellinvcplx} and $\gp_{1}\neq\g_{1}.$

        Proof of algebraic closure for the complex Cauchy
        sequences is limited to showing the existence of a
        Cauchy sequence whose square is $N$ equal to the constant
        sequence $\{|-,1\rangle\}_{c},$ (equivalent to a solution of
        $x^{2}=-1).$ This is trivial because the square of the
        constant sequence $\{|+,i1\rangle\}_{c}=
        \{\dd_{+,m}\bd_{1,m}|0\rangle\}_{c}$ equals $\{|-,1\rangle\}_{c}.$ Also the
        square of any Cauchy sequence $\{\psi_{n}\},$ that is $C$ equal to
        $\{|+,i1\rangle\}_{c},$ is $C$ equal to  $\{|-,1\rangle\}_{c}.$

        \subsection{Completeness}
        Another needed property of Cauchy sequences  is that of
        completeness.  This property is different from those
        discussed so far in that it deals with sets or sequences
        of Cauchy sequences of states
        in $\mathcal F.$ These have not been used so far in
        the development.  To this end it is useful to use a
        double indexing $|\g_{n,m}s_{n,m},(\g_{1})_{n,m}t_{n,m}\rangle$
        for complex string states. Here $m$ is the sequence index
        and $n$ labels the $nth$ component in the $mth$ sequence.

        To save on notation let $|\g_{n,m}s_{n,m},(\g_{1})_{n,m}t_{n,m}\rangle$
        be denoted by $|x_{n,m}\rangle.$ Also let
        $Re|x_{n,m}\rangle=|\g_{n,m}s_{n,m}\rangle$ and
        $Im|x_{n,m}\rangle=|(\g_{1})_{n,m}t_{n,m}\rangle.$ From the
        indexing one sees that $\{\{|x_{n,m}\rangle\}_{n}\}_{m}$
        denotes a double sequence of states where
        $\{|x_{n,m}\rangle\}_{n}$ is the $mth$ sequence and
        $|x_{n,m}\rangle$ is the $nth$ state in the $mth$
        sequence. For linear superposition states a similar
        representation of sequences of sequences is denoted by
        $\{\{\psi_{n,m}\}_{n}\}_{m}.$

        The proof of completeness requires showing that  every
        sequence of Cauchy sequences that is itself Cauchy,
        converges to a Cauchy sequence that is unique up to
        $=_{C}.$   There are two Cauchy conditions to consider,
        one for each sequence in the sequence and one for the
        sequence of sequences. A sequence
         $\{\{|x_{n,m}\rangle\}_{n}\}_{m}$ of Cauchy
         sequences is itself Cauchy if \begin{equation}\label{CauCau}
         \begin{array}{l}\mbox{For each
         $\ell$ there is an $h$ such that for all $j,k>h$ }\\
         |\{Re|x_{n,j}\rangle\}_{n}-_{R}\{Re| x_{n,k}\rangle\}_{n}|_{R}<_{R}
         \{|+,-\ell\rangle\}_{c}  \mbox{ and} \\
         |\{Im|x_{n,j}\rangle\}_{n}-_{I}\{Im| x_{n,k}\rangle\}_{n}|_{R}<_{R}
         \{|+,-\ell\rangle\}_{c}.\end{array}\end{equation}
         Here $\{|+,-\ell\rangle\}_{c}$ is the constant
         sequence of states $|+,-\ell\rangle,$ and $|\{Re|x_{n,j}
         \rangle\}_{n}-_{R}\{Re|x_{n,k}\rangle\}_{n}|_{R}$ and $|\{Im|x_{n,j}
         \rangle\}_{n}-_{I}\{Im|x_{n,k}\rangle\}_{n}|_{R}$ are the Cauchy
         state sequences that are the absolute values of the
         differences between the two real state Cauchy sequences
         $\{Re|x_{n,j}\rangle\}_{n}$ and $\{Re|x_{n,k}\rangle\}_{n}$
         and the two imaginary state Cauchy sequences
         $\{Im|x_{n,j}\rangle\}_{n}$ and $\{Im|x_{n,k}\rangle\}_{n}.$

         These two difference sequences are $R$  equal to the two
         sequence of states that are absolute values of the differences
         of the real part and of the imaginary parts of the component
         states:\footnote{That is, the absolute value of the difference
         of two Cauchy sequences is $R$ equal to the  sequence whose
         elements are the absolute values of the difference of the individual
         sequence elements.} \begin{equation} \label{difseq}\begin{array}{l}
         |\{Re|x_{n,j}\rangle\}_{n} -_{R}\{Re|x_{n,k}\rangle\}_{n}|_{R}\\
         \hspace{1cm}=_{R}\{|(|Re (x_{n,j})-_{R}Re (x_{n,k})|_{R})\rangle\}_{n}; \\
         |\{Im|x_{n,j}\rangle\}_{n} -_{I}\{Im|x_{n,k}\rangle\}_{n}|_{R} \\
         \hspace{1cm}=_{R} \{|(|Im (x_{n,j})-_{I}Im (x_{n,k})|_{R})
         \rangle\}_{n}.\end{array}\end{equation} Here $Re( x_{n,j})=
         \g_{n,j}s_{n,j}$ and $Im (x_{n,j})=(\g_{1})_{n,j}t_{n,j}.$
         Because of the substitution property of $=_{R},$ the righthand
         sequences in the above also satisfy the Cauchy conditions
         of Eq. \ref{CauCau}. The subscript $R$ on the absolute
         value of the difference of two imaginary state sequences
         accounts for the fact that absolute values of imaginary
         numbers are real.

         Convergence of a sequence $\{\{|x_{n,m}\rangle\}_{n}\}_{m}$ of
         Cauchy sequences to a sequence $\{|x^{\p}_{n}\rangle\}_{n}$ is
         expressed by \begin{equation}\label{Caucondri}
         \begin{array}{l}\mbox{For each $\ell$ there is an $h$ such
         that for all $j>h$ }\\
         |\{Re|x_{n,j}\rangle\}_{n}-_{R}\{Re|x^{\p}_{n}\rangle\}_{n}|_{R}<_{R}
         \{|+,-\ell\rangle\}_{c} \mbox{ and} \\ |\{Im|x_{n,j}\rangle\}_{n}-_{I}
         \{Im|x^{\p}_{n}\rangle\}_{n}|_{R}<_{R} \{|+,-\ell\rangle\}_{c}.
         \end{array}\end{equation}  Here the double inequality for
         the real and imaginary parts can be replaced by a single
         inequality:\begin{equation}\label{CauConv}
         |\{|x_{n,j}\rangle\}_{n}-_{C}\{|x^{\p}_{n}\rangle\}_{n}|_{R}<_{R}
         \{|+,-\ell\rangle\}_{c}.\end{equation}

         It should be emphasized that this
          definition of convergence is entirely different from
          those based on the usual properties of states and
          operators in ${\mathcal F}.$ The latter include
          definitions based on norm convergence of state sequences or on
          definitions of statistical distance between states
          \cite{Wootters,Braunstein,Matjey}.

          A complete proof of completeness will not be given here.
          Instead some salient aspects of a proof, which follows that in
          \cite{Hewitt} for equivalence classes of Cauchy sequences
          of rational numbers, will be outlined.

          To prove completeness one looks first at Cauchy sequences
          of constant sequences of string states $\{|\g_{n}s_{n},
          (\g_{1})_{n}t_{n}\rangle\}_{c}$ for each $n.$ The goal is to show that
          \begin{equation}\label{limconstseq}
          \lim_{n\rightarrow\infty} \{|\g_{n}s_{n},
          (\g_{1})_{n}t_{n}\rangle\}_{c}=_{C}\{|\g_{n}s_{n},
          (\g_{1})_{n}t_{n}\rangle\}_{n}.\end{equation}  To get this result, one
          notes that for any $k,$
          \begin{equation}\label{equivcomp}\begin{array}{l}
          \{|\g_{n}s_{n},(\g_{1})_{n}t_{n}\rangle\}_{n}-_{C}\{|\g_{k}s_{k},
          (\g_{1})_{k}t_{k}\rangle\}_{c} \\ \hspace{1cm}=_{C}\{|\g_{n}s_{n},
          (\g_{1})_{n}t_{n}-_{C}\g_{k}s_{k},(\g_{1})_{k}t_{k}\rangle\}_{n}.\end{array}
          \end{equation} From the Cauchy conditions for the
          sequence of constant sequences, \begin{equation}\label{Causeqconst}
          \begin{array}{c}|\{|\g_{j}s_{j}\rangle\}_{c}-_{R}\{|\g_{k}s_{k}
          \rangle\}_{c}|_{R} <_{R}\{|+,-\ell\rangle\}_{c} \\
          |\{|(\g_{1})_{j}t_{j}\rangle\}_{c}-_{I}
          \{|(\g_{1})_{k}t_{k}\rangle\}_{c}|_{R} <_{R}\{|+,-\ell\rangle\}_{c}
          \\ \mbox{for all $j,k,>h$},\end{array}\end{equation} one
          sees that both the real and imaginary components of the
          sequence on the right hand side of Eq. \ref{equivcomp}
          are $<_{A}|+,-\ell\rangle$ for all $n,k>$ some $h.$ Eq.
          \ref{limconstseq} follows from this.

          Similar arguments apply to more general Cauchy sequences
          of Cauchy sequences $\{\{|x_{n,f}\rangle\}_{n}\}_{f}$
          where $|x_{j,f}\rangle\neq_{A} |x_{k,f}\rangle$ is possible.
          Here the goal is to show that \begin{equation}\label{diaglim}
          \lim_{f\rightarrow\infty}\{|x_{n,f}\rangle\}_{n}=_{C}\{|x_{n,n}
          \rangle\}_{n}.\end{equation} where $\{|x_{n,n}\rangle\}_{n}$ is
          the desired limit $\{x'_{n}\rangle\}$ of Eq. \ref{Caucondri}.

          One starts by  noting thatthe Cauchy conditions for the sequence
          of sequences are\begin{equation}\label{CCdiag1}\begin{array}{c}|\{Re|x_{n,f}
         \rangle\}_{n}-\{Re|x_{n,g}\rangle\}_{n}|<_{R}\{|+,-\ell\rangle\}_{c}
          \\ |\{Im|x_{n,f}\rangle\}_{n}-\{Im|x_{n,g}\rangle\}_{n}|<_{R}
          \{|+,-\ell\rangle\}_{c} \\\mbox{for all $f,g>$ some $h$.}
          \end{array}\end{equation} Consider the sequence
          $|\{|x_{f,f}\rangle\}_{c}-\{|x_{n,f}\rangle\}_{n}
          |=_{R}\{|(|x_{f,f}-x_{n,f}|)\rangle\}_{n}.$
          Because each sequence, $\{|x_{n,f}\rangle\}_{n},$ is Cauchy,
         \begin{equation}\label{diagffnf}\begin{array}{c}
         \{|(|Re(x_{f,f}-_{R}x_{n,f})|_{R})\rangle\}_{n}<_{R}\{|+,-\ell\rangle\}_{c}
          \\ \{|(|Im(x_{f,f}-_{I}x_{n,f}) |_{R})\rangle\}_{n} <_{R}
          \{|+,-\ell\rangle\}_{c} \\ \mbox{for $f>$ some $h.$} \end{array}
         \end{equation} From this it follows that
         $\{|x_{n,n}\rangle\}_{n}$ is Cauchy and, by Eq.
         \ref{limconstseq},\begin{equation}\label{ffnn}
         \lim_{f\rightarrow\infty}\{|x_{f,f}\rangle\}_{c}
         =_{C}\{|x_{n,n}\rangle\}_{n}.\end{equation} Eq.
         \ref{diaglim} follows because, for sufficiently large $f$,
         \begin{equation}\label{nnfn}\begin{array}{l} |\{|x_{n,n}
         \rangle\}_{n}-_{C}\{|x_{,n,f}\rangle\}_{n}|_{R} \\ \hspace{0.5cm}<_{R}
        |\{|x_{n,n}\rangle\}_{n}-_{C}\{|x_{f,f}\rangle\}_{c}|_{R} \\
        \hspace{1cm}+_{R} |\{|x_{f,f}\rangle\}_{c}-_{C}\{|x_{n,f}
        \rangle\}_{n}|_{R}.\end{array}\end{equation}

          \section{Representation of Real and Complex Numbers in
          Quantum Theory}\label{RRCNQT}

          \subsection{Equivalence Classes of Cauchy Sequences}

          In the preceding it has been shown or made plausible that
          Cauchy sequences of states in $\mathcal F$ have properties
          corresponding to those of real and complex numbers.  If these
          properties can be lifted to equivalence classes of these sequences,
          then the sets of equivalence classes are real or complex numbers.

          To this end let $R^{\{\psi_{n}\}}$ and $C^{\{\psi_{n}\}}$ denote the
          sets of equivalence classes of Cauchy sequences based on
          real  and complex string state sequences. Two Cauchy sequences,
          $\{|\g_{n}s_{n},(\g_{1})_{n}t_{n}\rangle\}_{n}$ and
          $\{|\gp_{n}\sp_{n},(\gp_{1})_{n}\tp_{n}\rangle\}_{n},$ are
          equivalent or in the same equivalence class if and only
          if they are $C$
          equal:\begin{equation}\label{equiv=}\begin{array}{l}
          \{|\g_{n}s_{n},(\g_{1})_{n}t_{n}\rangle\}_{n}\equiv\{|\gp_{n}
          \sp_{n},(\gp_{1})_{n}\tp_{n}\rangle\}_{n} \\ \hspace{.5cm}
          \leftrightarrow\{|\g_{n}s_{n},(\g_{1})_{n}t_{n}\rangle\}_{n}=_{C}
          \{|\gp_{n}\sp_{n},(\gp_{1})_{n}\tp_{n}\rangle\}_{n}.\end{array}\end{equation}
          This definition extends to sequences of other types of
          states. Thus the Cauchy sequence $\{\psi_{n}\}\equiv
          \{\psi^{\p}_{n}\}_{n}$ or $\{\psi_{n}\}\equiv\{|\g_{n}s_{n},(\g_{1})_{n}
          t_{n}\rangle\}_{n}$ if and only if $\{\psi_{n}\}=_{C}
          \{\psi^{\p}_{n}\}_{n}$ or $\{\psi_{n}\}=_{C}\{|\g_{n}s_{n},(\g_{1})_{n}
          t_{n}\rangle\}_{n}.$ Similar relations hold for Cauchy
          sequences of real and imaginary string states and
          their superpositions. Thus  \begin{equation}\label{Requal}
          \begin{array}{l}\{|\g_{n}s_{n}\rangle\}_{n}\equiv
          \{|\gp_{n}\sp_{n}\rangle\}_{n} \\ \hspace{1cm}
          \leftrightarrow\{|\g_{n}s_{n}\rangle\}_{n}=_{R}
          \{|\gp_{n}\sp_{n}\rangle\}_{n},\end{array}\end{equation}with a
          similar relation for $=_{I}.$

         As was noted in the introduction, each equivalence class
         in $C^{\{\psi_{n}\}}$ and $R^{\{\psi_{n}\}}$ is larger than
         the corresponding class of classical Cauchy sequences of rational
         numbers. This occurs because Cauchy sequences of states $\{\psi_{n}\}$
         where each $\psi_{n}$ is a linear superposition of
          complex  string states have no classical counterparts.
          However expansion of the notion of convergent sequences from
          classical rational string numbers to quantum states does not create more
          equivalence classes. This is an immediate consequence of the
          proof,  Eqs. \ref{defpjkl}-\ref{P<Qri}, that for any Cauchy
          sequence $\{\psi_{n}\}$ of superposition states, there is
          a Cauchy sequence $\{|\g_{n}s_{n},(\g_{1})_{n}t_{n}\rangle\}_{n}$ of
          string states where $\{\psi_{n}\}=_{C}\{|\g_{n}s_{n},
          (\g_{1})_{n}t_{n}\rangle\}_{n}.$ A similar relation holds for Cauchy
          sequences of superpositions of real string states and of
          imaginary string states.

         This result is satisfying because it is consistent
         with the requirement that $C^{\{\psi_{n}\}}$ and $R^{\{\psi_{n}\}}$
         are isomorphic to the ground
         sets $R$ and $C$ that are the base for the Fock space
         $\mathcal F.$ An isomorphism, $M,$ from $C^{\{\psi_{n}\}}$ to $C$
         (and $R^{\{\psi_{n}\}}$ to $R$) can be easily constructed by
         use of the operator $\tilde{N}$ defined in Eq.
         \ref{def1N}. This will not be done here as it is
         straightforward and adds nothing to the development.

         It is often useful to let individual Cauchy sequences
         stand for  equivalence classes. This practice is
         often done and does no harm here. It also makes
         various aspects easier to deal with as one can
         work with individual Cauchy sequences rather than with
         equivalence classes.

         The proof that $R^{\{\psi_{n}\}}$ and $C^{\{\psi_{n}\}}$ are real
         and complex numbers requires that one lift the basic
         properties $=_{X},\leq_{X}$ and operations $\tilde{+}_{X},\tilde{-}_{X},$ and
         $\tilde{\times}_{X},\tilde{\div}_{X}$ on Cauchy sequences up to
         the equivalence classes.  To do this it is essential that the
         truth value of each property, and the result of each operation, is preserved
         under $=_{R}$(for real states), $=_{I}$ (for imaginary states), and
         $=_{C}$ (for complex states) based substitution. For instance, if
         ${\mathcal P}(\{|x_{n}\rangle\}_{n},\{|y_{n}\rangle\}_{n})$ is true for
         $\{|x_{n}\rangle\}_{n},\{|y_{n}\rangle\}_{n}$ and
         $\{\psi_{n}\}=_{C}\{|x_{n}\rangle\}_{n}$ and
         $\{\psi^{\p}_{n}\}_{n}=_{C}\{|y_{n}\rangle\}_{n},$ then
         $\mathcal{P}(\{\psi_{n}\},\{\psi^{\p}_{n}\}_{n})$ should be true.
         If $\tilde{O}(\{|x_{n}\rangle\}_{n},\{|y_{n}\rangle\}_{n})$
         represents an outcome sequence for the operation of
         $\tilde{O}$ on the sequence pair
         $\{|x_{n}\rangle\}_{n},\{|y_{n}\rangle\}_{n},$ then
         $\tilde{O}(\{|x_{n}\rangle\}_{n},\{|y_{n}\rangle\}_{n})=_{C}
         \tilde{O}(\{\psi_{n}\},\{\psi^{\p}_{n}\}_{n})$ should
         be true.

         An example of this invariance follows from the global
         definitions of the basic operations and relations. If
         $\{|\g_{n},s_{n};\g'_{n},t_{n}\}$ is a sequence of states
         at location $(m,h)$ and
         $\tilde{T}\{|\g_{n},s_{n};\g'_{n},t_{n}\}$ is a translation
         of the sequence to location $(m,h')$, then\begin{equation}
         \label{seqtrans}\tilde{T}\{|\g_{n},s_{n};\g'_{n},t_{n}\}
         =_{X}\{|\g_{n},s_{n};\g'_{n},t_{n}\}.\end{equation} Also
         the two sequences have exactly the same properties relative
         to $<_{X}$ and the basic operations.

         This raises the question of determining  which properties
         are preserved for $=_{R},$ $=_{I},$ or $=_{C}$ based substitution.
         At the least one would expect that all mathematical properties
         and operations described and used in the theory of complex
         analysis would be included.  One approach is to use the standard
         construction of terms and formulas in languages as described in
         mathematical logic \cite{Smullyan,Shoenfield}. Here the set of
         operations would be the smallest set that contains
         $\tilde{+},\tilde{\times},$ and their inverses, and is
         closed under all finite combinations of these operations
         and the taking of limits. The set of properties would be
          the smallest set that contains the basic relations
          $=_{R},<_{R}$ (or $=_{I},<_{I}$ or $=_{C},<_{C}$) and
          operations, and is closed under the use of logical connectives
          and existential quantifiers.

         Expansion of the above definitions to include more operations and
         properties may lead one to very difficult questions. These include
         defining the difference between physical and mathematical
         properties, and determining exactly what would be meant by requiring
         that "all" mathematical properties and operations should be
         included. These questions will not be dealt with here as they
         are outside the scope of this work.

          \section{Discussion}
        The goal of this paper has been to show that the sets, $R^{\{\psi_{n}\}},
        C^{\{\psi_{n}\}},$ of equivalence classes of Cauchy sequences
        of real and complex string states states and their superpositions
        satisfy the required properties of real and complex numbers. This
        was done with no reference to the properties of numbers in the
        underlying $R$ and $C$ that are
        the base of $\mathcal F.$ This includes the definition of the
        Cauchy condition, and definitions and properties of the relations
        $=_{R},=_{I},=_{C},<_{R},<_{I},$ and $<_{C},$
        and the addition, subtraction, multiplication, and division
        operations. For sequences of superposition states $\{\psi_{n}\}$
        this is not possible because the coefficients of the
        $\psi_{n}$ are complex numbers in $C$.

        It was also noted that the equivalence classes of Cauchy
        sequences in $C^{\{\psi_{n}\}}$ and $R^{\{\psi_{n}\}}$ are larger than the
        corresponding equivalence classes in any classical $R$ and
        $C$. This follows from the existence of Cauchy sequences
        of quantum states that have no classical equivalences. These
        states can be somewhat counterintuitive.  For example,
        following Eqs. \ref{exam1} and \ref{defBn}, let $s$ be a
        $0-1$ function with domain $[0,-\infty]$ and $t$ be the
        constant $0$ function on the same domain. Define $\psi_{n}$
        by\begin{equation}\label{example}\begin{array}{l}
        \psi_{n}=\cd_{+,m,h}(\ad)^{s_{[0,-n+1]}}\dd_{+,m,h}(\bd)^{t_{[0,-n+1]}}\\
        \hspace{1cm}\times(\frac{1}{\sqrt{2}}(\ad_{1,-n}\bd_{0,-n}+\ad_{0,-n}
        \bd_{1,n})|0\rangle\end{array}
        \end{equation} where\begin{equation}\label{adsbdt}\begin{array}{l}
        (\ad)^{s_{[0,-n+1]}}=\ad_{s(0),0}\ad_{s(-1),-1}\cdots\ad_{s(-n+1),-n+1}\\
        (\bd)^{t_{[0,-n+1]}}=\bd_{t(0),0}\bd_{t(-1),-1}\cdots\bd_{t(-n+1),-n+1}.
        \end{array}\end{equation} Each state in this sequence is an entangled
        state of real and imaginary components. However, as a Cauchy
        sequence, $\{\psi_{n}\}$ is a real number with no imaginary component.

        This, and other examples, show that the quantum equivalence classes are
        larger than the classical ones, however, no new classes are created. It follows that
        $R^{\{\psi_{n}\}}$ and $C^{\{\psi_{n}\}}$ are isomorphic to $R$ and $C$.
        This shows that $C^{\{\psi_{n}\}}$ and $R^{\{\psi_{n}\}}$ are in every way
        just as good and valid a representation of  real and complex
        numbers as are the original $R$ and $C$ over which the Fock space
        $\mathcal F$ was constructed.

        In this case there is no reason why one could not use $C^{\{\psi_{n}\}}$
        and $R^{\{\psi_{n}\}}$ to be the base of physical theories such as QED,
        QCD, special and general relativity, string theory, etc.  Also
        $(R^{\{\psi_{n}\}})^{4}$ is just as good a representation of a space time
        manifold as is $R^{4}$. This raises all sorts of interesting open
        questions concerning the relations between states and properties
        of the physical systems described by the states in $\mathcal F,$  and
        those described by the theories based on  $R^{\{\psi_{n}\}}$ and
        $C^{\{\psi_{n}\}}.$

       One interesting question is based on the observation that a Fock
        space  $\mathcal F$ of states of finite qubit strings, equipped
        with an associated Hamiltonian $H$ and a discrete space time
        lattice $\mathcal L$ of points in $R^{4},$ can be used to describe
        the lattice quantum dynamics of the qubit strings. As noted
        $\mathcal F$ is based on $R$ and $C$. Let ${\mathcal F}^{\{\psi_{n}\}}$
        be another Fock space based on the  real and complex numbers in
        $R^{\{\psi_{n}\}}$ and $C^{\{\psi_{n}\}}$ which consist of Cauchy
        sequences of the string states or their linear superpositions in
        $\mathcal F.$

        Now consider the situation where ${\mathcal F}^{\{\psi_{n}\}}$ with
        some Hamiltonian $H^{\prime}$ and corresponding space time
        lattice $\mathcal{L}^{\{\psi_{n}\}}$ of points in
        $(R^{\{\psi_{n}\}})^{4}$ describes the quantum dynamics of the
        \emph{same} physical systems whose quantum dynamics is
        described by $\mathcal F,$ $H,$ and $\mathcal{L}.$ This is an
        interesting situation since the states in $\mathcal F$ of the
        physical systems are states in the Cauchy sequences in $C^{\{\psi_{n}\}}$
        on which ${\mathcal F}^{\{\psi_{n}\}}$ and the space time
        $(R^{\{\psi_{n}\}})^{4}$ are based. Is this situation even possible? Do
        there exist physical systems whose quantum dynamics is described
        by both $\mathcal F,$  $\mathcal{L}$ in $R^{4}$ and $H$, and by ${\mathcal
        F}^{\{\psi_{n}\}},$  $\mathcal{L}^{\{\psi_{n}\}}$ in
        $(R^{\{\psi_{n}\}})^{4},$ and $H^{\prime}$? It is hoped to
        investigate this and related questions in future work.

        This all suggests that this process can be iterated,
        leading to a hierarchy of Fock spaces over complex numbers
        as equivalence classes of Cauchy sequences of states that
        are based on the previous space and complex numbers in the
        iteration \cite{BenFIQRF}.  The existence of such a hierarchy
        suggests that it may be of interest to study the relationship
        between two neighboring spaces and states in the iteration.

        Of particular interest is the relation between the
        original $R$ and $C,$ and $R^{\{\psi_{n}\}}$ and $C^{\{\psi_{n}\}}.$ Here,
        as in other physical theories $R$ and $C$ is taken for
        granted, or as given, without thought as to what the
        structure is, if any, of the numbers in $R$ and $C.$  If they are
        equivalence classes of Cauchy sequences of complex rational
        numbers, then what are the rational numbers?  Pursuing
        this line leads back to the natural numbers or nonnegative
        integers.   Depending on ones point of view they can
        either be accepted as primary and unanalyzable, or one can
        ask what they are and how they relate to physical systems
        and quantities.\footnote{In the description given here natural
        number states have the form $|\a s\rangle=\cd_{+,m,h}\ads|0\rangle$
        where $s(j,h)=1\rightarrow j\geq m.$} This is part of the
        more general question of the foundational relation between
        mathematics and physics \cite{BenTCTPM,Tegmark2}.

       It should be noted that for any function or property on
       $R$ or $C,$ there is a corresponding function or property
       on $R^{\{\psi_{n}\}}$ or $C^{\{\psi_{n}\}}.$ For example, corresponding to a
       metric on $R$ one has a metric on $R^{\{\psi_{n}\}}$ defined by
        \begin{equation}\label{distance}\begin{array}{l}
        D(\{|\g_{n}s_{n}\rangle\}_{n},\{|\gp_{n}\sp_{n}\rangle\}_{n})
         \\ \hspace{1cm}=|\{|\g_{n}s_{n}\rangle\}_{n}-_{R}\{|\gp_{n}
         \sp_{n}\rangle\}_{n}|_{R}
        \\ \hspace{1.5cm} =_{R} \{|(|\g_{n}s_{n}-_{A}\gp_{n}\sp_{n}|_{R})
        \rangle\}_{n}.\end{array}\end{equation}  The right hand
        term in the above is a Cauchy sequence whose elements are
        the states that are the absolute values of the differences between
        $|\g_{n}s_{n}\rangle$ and $|\gp_{n}\sp_{n}\rangle$ for
        $n=1,2,\cdots.$  A similar map can be given for $C^{\{\psi_{n}\}}$
        by replacing $|\g_{n}s_{n}\rangle$ with
        $|\g_{n}s_{n},(\g_{1})_{n}t_{n}\rangle,$ etc. This shows that
        $R^{\{\psi_{n}\}}$ and $C^{\{\psi_{n}\}}$ (and $R$ and $C$) are metric spaces
        \cite{Randolph,Rudin}.

        It is clear that there is much to do. Future work includes
        more examination of the iterative hierarchy noted above. Also the
        treatment should be expanded to include qukits for any base
        $k\geq 2,$ not just $k=2.$

          \section*{Acknowledgements}
          This paper has greatly benefitted from a critical
          reading by Fritz Coester.
          This work was supported by the U.S. Department of Energy,
          Office of Nuclear Physics, under Contract No. W-31-109-ENG-38.


\begin{thebibliography}{99}

            \bibitem{Wigner}
            E. Wigner, Commun. Pure, Applied Mathematics,
            \textbf{13}, No. 1, (1960).

            \bibitem{Hamming}
            R. W. Hamming, Amer. Math. Monthly \textbf{87} No. 2,
            (1980).

            \bibitem{Davies}
            P. C. W. Davies, Arxiv Preprint quant-ph/0703041.

            \bibitem{Tegmark1}
            M. Tegmark, Ann. Phys. \textbf{270}, 1 (1998) (Arxiv
            preprint gr-qc/9704009).

            \bibitem{Tegmark2}
            M. Tegmark, Arxiv preprint 0704.0646v1 [gr-qc].

            \bibitem{Schmidhuber}
            J. Schmidhuber, Arxiv preprint quant-ph/0011122v2.

            \bibitem{Weinberg}
            S. Weinberg, \emph{Dreams of a Final Theory},
            Vintage Books, New York, 1994

             \bibitem{BenTCTPM}
            P. Benioff, \textit{Foundations of Physics},
            \textbf{32}, 989-1029, (2002)  (Arxiv preprint
            quant-ph/0201093); \textit{Foundations of Physics},
            \textbf{35}, 1825-1856, (2005)
             (Arxiv preprint quant-ph/0403209).

              \bibitem{Isham}
            C. J. Isham, Arxiv Preprint, quant-ph/0206090.

            \bibitem{Burkill}
            J. C. Burkhill and H. Burkill, \emph{A Second Course
            in Mathematical Analysis}, Cambridge University Press,
            Cambridge, England, 1970.

            \bibitem{Litvinov}
            G. L. Litvinov, V. P. Maslov, and G. B. Shpiz,
            Archives preprint, quant-ph/9904025, v5, 2002.

            \bibitem{Corbett}
            J. V. Corbett and T. Durt, Archives preprint,
            quant-ph/0211180 v1 2002.

            \bibitem{Tokuo}
            K. Tokuo, \emph{Int. Jour. Theoretical Phys.},
            \textbf{43}, 2461-2481, 2004.

            \bibitem{Takeuti}
            G. Takeuti, \emph{Two Applications of Logic in
            Mathematics}, Princeton University Press, New Jersey,
            1978.

            \bibitem{Davis}
            M. Davis, \emph{Int. Jour. Theoretical Phys.},
            \textbf{16}, 867-874, (1977).

            \bibitem{Titani}
            S. Titani and H. Kozawa, Int. Jour. Theoret. Physics
            \textbf{42}, 2575-2602, (2003).

            \bibitem{Finkelstein}
            D. Finkelstein, \emph{Quantum Relativity}, Springer-Verlag,
            Heidelberg (1996)

            \bibitem{Schlesinger}
            K. G. Schlesinger, Jour. Math. Phys. \textbf{40},
            1344-1358, (1999).

            \bibitem{Krol}
            Jerzey Krol, "A Model of Spacetime. The Role of
            Interpretations in Some Grothendieck Topoi", preprint,
            (2006).

            \bibitem{Wootters1}
            W.K. Wootters and W.H. Zurek,  Nature \textbf{299}, 802-803,
            (1982).

             \bibitem{BenRCRNQM}
            P. Benioff, Phys. Rev. \textbf{A72}, 032314
            (2005); Int. Jour. Modern Phys. B \textbf{70},
            1730-1741, (2006) (Arxiv preprint quant-ph/0508038).

              \bibitem{Hewitt}
             E. Hewitt and K. Stromberg, \emph{Real and Abstract
             Analysis}, Springer-Verlag New York, Inc. 1965.

             \bibitem{Randolph}
            J. Randolph, \emph{Basic Real and Abstract Analysis},
            Academic Press, New York, 1968.

            \bibitem{Wootters}
             W. K. Wootters, \emph{Phys. Rev. D}, \textbf{23}, 357-362 (1981).

             \bibitem{Braunstein}
             S. L. Braunstein and C. M. Caves,  \emph{Phys.Rev. Letters} \textbf{72},
             3439-3442 (1994).

             \bibitem{Matjey}
             A. Matjey, P. W. Lambert, and A. Plastino, \emph{European Physics Journal
             D}, \textbf{32}, 413-419 (2005).

            \bibitem{Smullyan}
            R. Smullyan, \emph{G\"{o}del's Incompleteness
            Theorems}, Oxford University Press, Oxford, 1992.

            \bibitem{Shoenfield}
            J. R. Shoenfield, \emph{Mathematical Logic}, Addison
            Weseley, Reading, Ma. 1967.

             \bibitem{BenFIQRF}
            P. Benioff, Arxiv preprint quant-ph/0604135.

            \bibitem{Rudin}
            W. Rudin, \emph{Principles of Mathematical Analysis}
            3rd Edition, McGraw Hill Book Co., New York, 1976.

             \end{thebibliography}
          \end{document}